\documentclass[sigconf,authorversion,natbib=false]{acmart}
\usepackage{amssymb}
\usepackage{multirow}
\usepackage{multicol}
\usepackage{listings}
\usepackage{adjustbox}
\usepackage{makecell}
\usepackage{enumitem}
\usepackage{colortbl}
\usepackage{booktabs}
\usepackage{listings}
\definecolor{ForestGreen}{RGB}{34,139,34}
\usepackage{algorithm}
\usepackage[noend]{algpseudocode}

\lstdefinestyle{interfaces}{
  float=tp,
  floatplacement=tbp,
}

\usepackage{listings}
\definecolor{dkgreen}{rgb}{0,0.6,0}
\definecolor{gray}{rgb}{0.5,0.5,0.5}
\definecolor{mauve}{rgb}{0.58,0,0.82}

\AtBeginDocument{%
  \providecommand\BibTeX{{%
    \normalfont B\kern-0.5em{\scshape i\kern-0.25em b}\kern-0.8em\TeX}}}
    
\copyrightyear{2023}
\acmYear{2023}
\setcopyright{rightsretained}
\acmConference[FPGA '23]{Proceedings of the 2023 ACM/SIGDA International Symposium on Field Programmable Gate Arrays}{February 12--14, 2023}{Monterey, CA, USA} \acmBooktitle{Proceedings of the 2023 ACM/SIGDA International Symposium on Field Programmable Gate Arrays (FPGA '23), February 12--14, 2023, Monterey, CA, USA} \acmDOI{10.1145/3543622.3573210} \acmISBN{978-1-4503-9417-8/23/02}

\begin{document}
\title{CHARM: \underline{C}omposing \underline{H}eterogeneous \underline{A}ccele\underline{R}ators
for \underline{M}atrix Multiply on Versal ACAP Architecture
}

\settopmatter{authorsperrow=5}
\author{Jinming Zhuang}
\affiliation{%
  \institution{\small{University of Pittsburgh} 
  \country{USA}}}
\email{jinming.zhuang@pitt.edu}

\author{Jason Lau}
\affiliation{%
  \institution{\small{University of California, Los Angeles} \country{USA}}}
\email{lau@cs.ucla.edu}

\author{Hanchen Ye}
\affiliation{%
  \institution{\small{University of Illinois at Urbana-Champaign} \country{USA}}}
\email{hanchen8@illinois.edu}

\author{Zhuoping Yang}
\affiliation{%
  \institution{\small{University of Pittsburgh} \country{USA}}}
\email{zhuoping.yang@pitt.edu}

\author{Yubo Du}
\affiliation{%
  \institution{\small{University of Pittsburgh} \country{USA}}}
\email{yubo.du@pitt.edu}

\author{Jack Lo}
\affiliation{%
  \institution{\small{Advanced Micro \\ Devices Inc.} \country{USA}}}
\email{jack.lo@amd.com}

\author{Kristof Denolf}
\affiliation{%
  \institution{\small{Advanced Micro \\ Devices Inc.} \country{USA}}}
\email{kristof.denolf@amd.com}

\author{Stephen Neuendorffer}
\affiliation{%
  \institution{\small{Advanced Micro \\ Devices Inc.} \country{USA}}}
\email{stephen.neuendorffer@amd.com}

\author{Alex Jones}
\affiliation{%
  \institution{\small{University of Pittsburgh} \country{USA}}}
\email{akjones@pitt.edu}

\author{Jingtong Hu}
\affiliation{%
  \institution{\small{University of Pittsburgh} \country{USA}}}
\email{jthu@pitt.edu}

\author{Deming Chen}
\affiliation{%
  \institution{\small{University of Illinois at Urbana-Champaign}\country{USA}}}
\email{dchen@illinois.edu}

\author{Jason Cong}
\affiliation{%
  \institution{\small{University of California, Los Angeles} \country{USA}}}
\email{cong@cs.ucla.edu}

\author{Peipei Zhou}
\affiliation{%
  \institution{\small{University of Pittsburgh}
  \country{USA}}}
\email{peipei.zhou@pitt.edu}

\renewcommand{\shortauthors}{Jinming Zhuang et al.} 

\begin{abstract}
Dense matrix multiply (MM) serves as one of the most heavily used kernels in deep learning applications. 
To cope with the high computation demands of these applications,
heterogeneous architectures featuring both FPGA and dedicated ASIC accelerators have emerged as promising platforms. 
For example, the AMD/Xilinx Versal ACAP architecture combines general-purpose CPU cores and programmable logic with AI Engine processors optimized for AI/ML.  An array of 400 AI Engine processors executing at 1~GHz can provide up to 6.4~TFLOPs performance for 32-bit floating-point (fp32) data. However, machine learning models often contain both large and small MM operations.  While large MM operations can be parallelized efficiently across many cores, small MM operations typically cannot.  We observe that executing some small MM layers from the BERT natural language processing
model on a large, monolithic MM accelerator in Versal ACAP achieved less than 5\% of the theoretical peak performance.
Therefore, one key question arises:\emph{ How can we design accelerators to fully use the abundant computation resources under limited communication bandwidth for end-to-end applications with multiple MM layers of diverse sizes?}

We identify the biggest system throughput bottleneck resulting from the mismatch of massive computation resources of one monolithic accelerator and the various MM layers of small sizes in the application. To resolve this problem, we propose the CHARM framework to compose \textbf{multiple diverse MM accelerator architectures} working concurrently towards different layers within one application. 
CHARM includes analytical models which guide design space exploration to determine accelerator partitions and layer scheduling.
To facilitate the system designs, CHARM automatically generates code, enabling thorough onboard design verification. 
We deploy the CHARM framework on four different deep learning applications, including BERT, ViT, NCF, MLP, on the AMD/Xilinx Versal ACAP VCK190 evaluation board.
Our experiments show that we achieve 1.46 TFLOPs, 1.61 TFLOPs, 1.74 TFLOPs, and 2.94 TFLOPs inference throughput for BERT, ViT, NCF, MLP, respectively, which obtain 5.29$\times$, 32.51$\times$, 1.00$\times$ and 1.00$\times$ throughput gains compared to one monolithic accelerator.
\end{abstract}

\ccsdesc[500]{Computer systems organization~Heterogeneous (hybrid) systems}
\ccsdesc[500]{Hardware~Hardware-software codesign}

\vspace{-8pt}
\keywords{Heterogeneous Architecture, Domain-Specific Accelerator, Versal ACAP, Mapping Framework, Matrix-Multiply, Deep Learning}

\maketitle

{
\vspace{-5pt}
\small
\textbf{ACM Reference Format:}

\noindent
Jinming Zhuang, Jason Lau, Hanchen Ye, Zhuoping Yang, Yubo Du, Jack Lo, Kristof Denolf, Stephen Neuendorffer, Alex Jones, Jingtong Hu, Deming Chen, Jason Cong, Peipei Zhou. 2023. CHARM: Composing Heterogeneous Accelerators for Matrix Multiply on Versal ACAP Architecture. In \textit{Proceedings of the 2023 ACM/SIGDA International Symposium on Field Programmable Gate Arrays (FPGA ’23), February 12–14, 2023, Monterey, CA, USA}. ACM, New York, NY, USA, 12 pages. https://doi.org/10.1145/3543622.3573210
}

\vspace{-5pt}
\section{Introduction}
\label{sec:intro}
Dense matrix multiply (MM) serves as one of the most heavily used kernels in many deep learning workloads, including BERT~\cite{vaswani2017attention} 
for natural language processing, NCF~\cite{he2017neural} for recommendations, ViT~\cite{dosovitskiy2020image} for vision classification, and MLP~\cite{wang2019benchmarking} for multilayer perceptron classification or regression. 
According to profiling results from Google~\cite{jouppi2017datacenter}, dense matrix multiply tasks occupied  90\% of Neural Network~(NN) inference workload in Google's data center in 2017. 
The increasing complexity of these applications leads to extreme demands for computation and data movement. 

\begin{figure}
    \centering
    \includegraphics[width=1.0\linewidth]{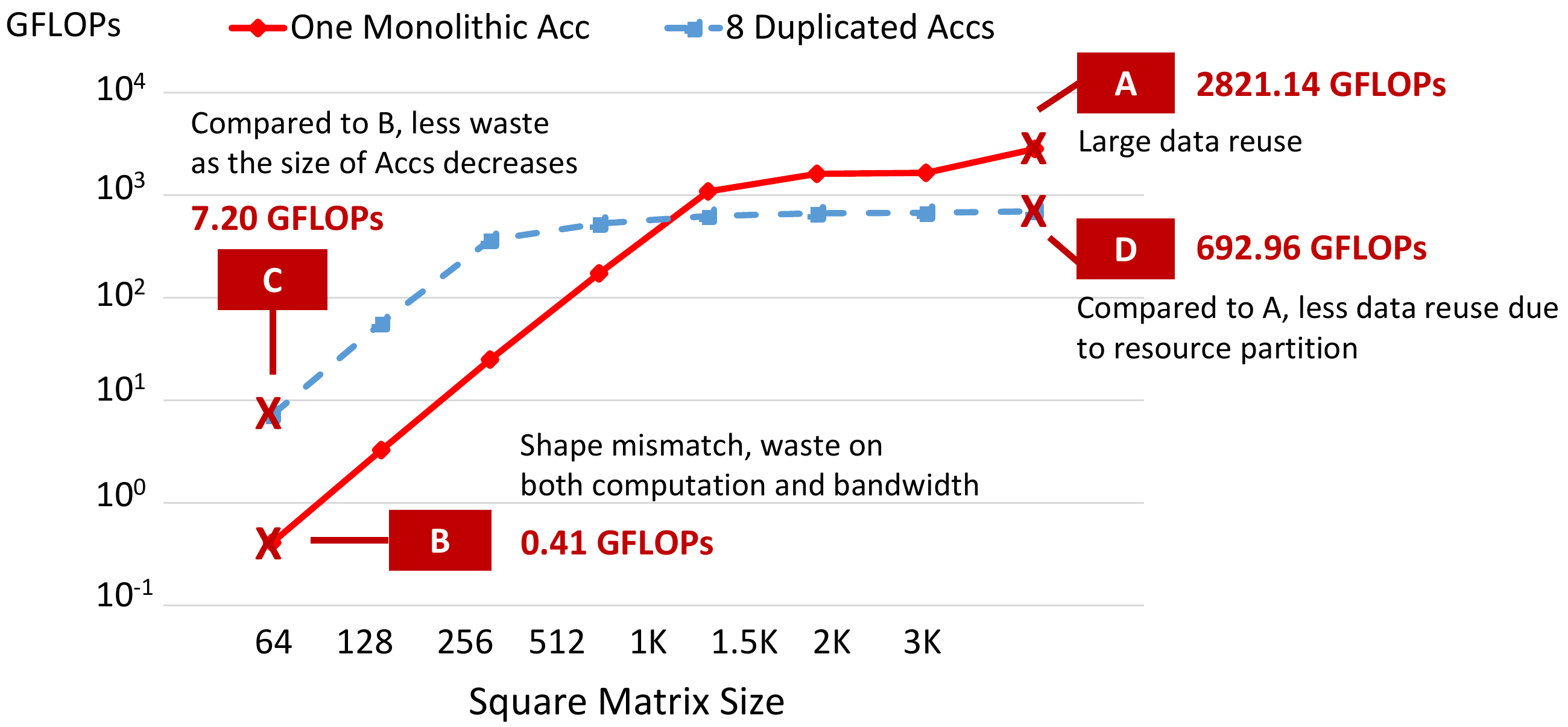}
    \vspace{-20pt}
    \caption{Throughput of square MM under different sizes.}
    \label{Throughput_size}
\end{figure}

\begin{figure}
    \centering
    \includegraphics[width=1\linewidth]{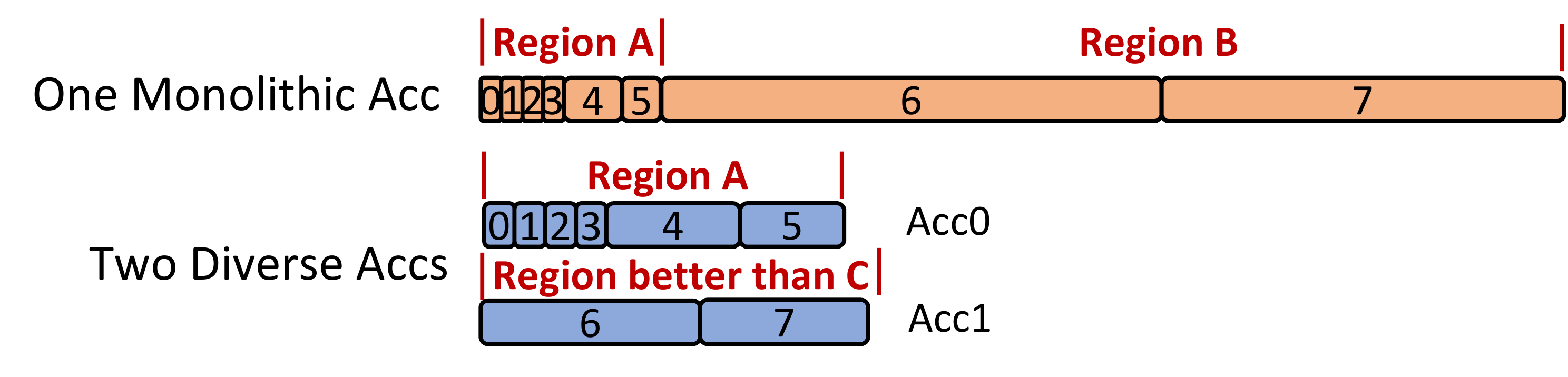}
    \caption{Execution timeline of one monolithic MM design vs. two diverse MM accs design for BERT on VCK190.}
    \label{fig:Motivation two}
\end{figure}

According to ~\cite{mutlu2023modern,oliveira2021damov,hassan2019crow,demmel2012communication}, the off-chip bandwidth has been a bottleneck for both the performance and energy efficiency of a system and a common trend on current platforms is that the off-chip bandwidth does not scale as fast as the computation resources.
Therefore, the first research question arises: 
\emph{How to sustain the faster scaling computation with the slower scaling off-chip bandwidth?}

A common solution is to increase data reuse by allocating more on-chip storage within an accelerator (acc).
As shown in asymptotic analysis in ~\cite{demmel2012communication}, the total off-chip communication volume in MM scales as $\mathcal{O}(\frac{1}{\sqrt{M}})$ where M is the on-chip tile size. If we increase the tile size, we can reduce the total communication volume, therefore reducing the pressure on the off-chip bandwidth.

In this work, we target on the AMD/Xilinx Versal ACAP architecture~\cite{Versal_ACAP}, which combines general-purpose CPU cores and programmable logic (PL) with AI Engine processors (AIE) optimized for AI/ML computation.  For example, we implemented an MM accelerator on an AMD/Xilinx VCK190 board using 384 AIEs and over 80\% on-chip URAM and BRAM resources. The red line in Figure ~\ref{Throughput_size} illustrates the performance of this accelerator.  
This design operates on a native tile size of 1536$\times$128$\times$1024 and achieves 2.8 TFLOPs throughput when carrying a tiled execution of a large square MM (point A).
However, when simply mapping different sizes of MM to such a design, the performance decreases significantly as the square MM size drops below 512, since each tile is padded to the native tile size of the accelerator.  For instance, at point B, the performance of such a monolithic design goes to 0.41~GFLOPs, which is 6880$\times$ lower than point A. 
Although padding is a common and simple approach to implementing small MM operations on a large accelerator, padding can waste both computation and bandwidth.

An alternative to padding is implementing multiple accelerators with smaller native tile sizes, potentially executing different tasks on each accelerator in parallel~\cite{VersalDPU}.
We apply this approach using eight independent accelerators with a native tile size of 256$\times$128$\times$256, illustrated by the blue dash line in Figure~\ref{Throughput_size}.   For small square MM operations with size 64, this approach achieves 7.2 GFLOPS at point C, approximately 17$\times$ speedup compared to point B. 

However, the smaller accelerator size also means less data reuse for large MM, with total throughput almost saturation when the operation size is larger than 256. 
When the MM size is 3072 (point D), the total throughput from eight duplicate accs is 4.08$\times$ smaller than point A in one monolithic design. 

These experiments expose two conflicting design goals.  Firstly, we want to implement large MM operations with sufficient data reuse to achieve the highest possible performance on the devices.  Secondly, we want to implement small MM operations while minimizing computation and communication overheads.  Neither of these simple designs seems able to achieve these design goals simultaneously.  
Therefore, the second research question arises: \emph{How to trade-off between the two design goals for real-world, end-to-end applications where MM layers with large and small sizes coexist?}

To illustrate how these conflicting design goals can affect the performance of practical machine learning models, we consider BERT~\cite{vaswani2017attention} as a representative workload containing MM layers with both large and small sizes.
In a transformer layer of BERT, there are a total of 8 types of MM kernels where Kernels 0-5 are large MMs and Kernels 6 and 7 are batch dots, i.e., small MMs. 
The detailed shapes can be referred to Table~\ref{tbl:Evaluation}.
Take Kernel 5 and Kernel 6 as examples,
Kernel 5 is an MM with the shape 3072$\times$1024$\times$4096,
Kernel 6 is a batch dot with the shape 96$\times$512$\times$512$\times$64, which means there are 96 small independent MMs sized at 512$\times$512$\times$64.

As shown in Figure~\ref{fig:Motivation two}, when using one monolithic MM accelerator,
Kernels 0-5 consume 92\% of the total BERT MM computation operations and 12\% of the total MM acc time. 
In contrast, Kernels 6-7 consume 8\% of the total operations but take 88\% of the total MM acc time. 
For Kernels 0-5, they lie in Region A (a region that performs similarly to Point A in Figure ~\ref{Throughput_size}), where the throughput of acc is more than 2082 GFLOPS.
For Kernel 6-7, they lie in Region B, where the throughput of acc is only 23.6 GFLOPS. 
Given there is a large portion of acc execution underutilized in the timeline, the overall MM acc throughput is only 276 GFLOPS.
Can we achieve a design for BERT that lies in region A, i.e., good for large MMs, and also in a region better than point C, i.e., good for small MMs with less or no waste computation/bandwidth?

Our answer is ``Yes". 
The key idea is to allocate more portion of the resources to accs dedicated to computing larger MMs and a smaller portion of the resources to other accs to compute smaller MMs at the same time, as shown in Figure~\ref{fig:Motivation two} where a two-diverse accs system is illustrated.
To achieve our design goals, we need to solve these \emph{new challenges}.
\textbf{First}, we need to achieve high computation utilization for every single acc, i.e., use the smaller acc(s) to reduce the waste for small MMs and use the larger acc(s) to maximize the data reuse for large MMs. 
\textbf{Second}, to maximize overall utilization while maintaining high throughput and low latency, we need to carefully overlap the execution time for these accs by co-optimizing workload and resource partitioning.
\textbf{Third}, to facilitate the design space explorations (DSE), we need analytical models to optimize the overall throughput under resource and bandwidth constraints. 
\textbf{Fourth}, to reduce the programming efforts for the system implementation, we need automatic code generation.
\textbf{Fifth}, to resolve the dependency of the kernels within the application graph when running multiple accs  we need an accelerator runtime to schedule kernels from different tasks onto the accs.

To answer the research questions, we propose the CHARM architecture and its corresponding automation framework, the CHARM framework.
Our contributions are summarized below:
\vspace{-5pt}
\begin{itemize}[leftmargin=*]
\item \textbf{CHARM Systematical Design Methodology on Versal}:
To achieve high computation and communication efficiency of each acc, in Section 4, we propose a thorough design methodology on Versal heterogeneous platform. We further provide automatic CHARM DSE (CDSE) to find the optimized single acc configuration. To the best of our knowledge, this is the first work that provides a detailed analysis of the systematical data movement and computation on Versal.

\item \textbf{CHARM Architecture and Framework:}
To achieve the design goals of good performance for MMs with both small and large sizes in an application, in Section~\ref{sec:dse_new}, we propose the CHARM architecture and the CHARM framework to find the optimized design.
In the CHARM framework, there are several modules. 
First, on top of CDSE, we propose the CHARM diverse accelerator composer (CDAC), which features a sort-based two-step search algorithm to find an optimized CHARM design in the polynomial time complexity instead of exponential time complexity.  
Furthermore, to automate the system implementation, CHARM automatic code generation (CACG) is proposed to generate source code files for AIEs, PL, and host CPU.
Lastly, the CHARM runtime system (CRTS) is launched in the host CPU that schedules different kernels to the accs for optimizing both task latency and overall system throughput.

\item We deploy the CHARM framework to accelerate four applications on VCK190 in Section~\ref{sec:experiments}.
Our on-board experiments demonstrate that CHARM achieves 1.46 TFLOPs, 1.61 TFLOPs, 1.74 TFLOPs, and 2.94 TFLOPs
inference throughput for BERT, ViT, NCF, MLP, respectively, which obtain 5.29$\times$, 32.51$\times$, 1.00$\times$, and 1.00$\times$, throughput gains compared to one monolithic accelerator.

\item \textbf{White-Box Open-Source Tools for Versal.} While AMD provides users a block-box IP for NN applications called DPU~\cite{VersalDPU}, we open-sourced our tools completely as a white-box with a detailed step-by-step guide to reproduce all of the results presented in this paper and for the other users to learn and leverage in their end-to-end systems. 
(\textbf{\emph{\url{https://github.com/arc-research-lab/CHARM}}})
\end{itemize}

\vspace{-10pt}
\section{Prior Work}
%
\label{sec:prior_work}
To achieve high throughput and energy efficiency, NN accelerators usually employ a large number of processing elements (PE) and share a similar memory hierarchy. That is, while the big bulk of data is stored in the off-chip memory, there are multiple levels of on-chip buffers, including the local memory attached to each PE and global shared memory, to further reduce the costly data movement from/to off-chip memory. Several works contribute to NN accelerators by discussing the data reuse opportunities, computation parallelism, and the choice of dataflow.

However, many of the prior works apply a one-size-fits-all monolithic design that cannot efficiently handle layers with huge differences in shapes and sizes (Eyeriss~\cite{chen2016eyeriss,chen2019eyeriss}, ShiDiannao~\cite{du2015shidiannao}, NPU~\cite{nurvitadhi2019compete,boutros2020beyond,fowers2018configurable} and others ~\cite{de2020fblas,de2020flexible,zhang2015optimizing,Moss_Krishnan_Nurvitadhi_Ratuszniak_Johnson_Sim_Mishra_Marr_Subhaschandra_Leong_2018}). AutoSA ~\cite{Wang_Guo_Cong_2021} is a polyhedral-based compilation framework that generates monolithic systolic array designs for dense matrices. Sextans and Serpens~\cite{song2022sextans,song2022serpens} are general-purpose monolithic accelerators for sparse matrices.
\cite{fccm18latte,fccm16} analyze layout and pipeline efficiency.
Other works like AMD DPU~\cite{VersalDPU}, Mocha~\cite{fpga21mocha} explore task-level parallelism by allocating multiple duplicate accs on the device without specializing each acc.
DNNBuilder \cite{zhang2018dnnbuilder} designs a dedicated acc for each layer according to the number of operations within the layer. DNNExplorer~\cite{zhang2020dnnexplorer} enhances DNNBuilder by combining dedicated accs for the first several layers and a monolithic acc for the rest of the layers.
While it employs multiple accelerators, it lacks a comprehensive exploration of workload assignments.
TETRIS~\cite{gao2017tetris} and TANGRAM~\cite{gao2019tangram} propose multiple dataflow optimizations within and across the NN layers to improve performance and energy efficiency. Although they offer diverse accelerator designs, they lack the DSE and workload assignment for high overall throughput. 
Herald~\cite{kwon2021heterogeneous} proposes an architecture with multiple diverse accelerators and explores the workload assignment and resource partition. Still, they choose several existing acc designs from their candidate pool, e.g., ShiDiannao~\cite{du2015shidiannao}, NVDLA~\cite{NVIDIA} without doing DSE for each acc.
FPCA~\cite{FPCA14fccm} and CHARM\textquotesingle12~\cite{charm12islped} propose a fully pipelined and dynamically composable coarse-grained reconfigurable architecture and compose loosely coupled accelerators for different kernels within an application via permutation network, which costs high in chip area.

In conclusion, we summarize the differences between our work and prior works in Table~\ref{Prior_Comparison}. 
Our work is capable of choosing the design from one monolithic, multiple duplicates, and multiple diverse accelerators, and each accelerator is a specialized design considering the different workload assignments, dataflow, and data parallelism strategies that are covered by our DSE.

\begin{table}
    \centering
\caption{\textcolor{black}{Comparison with prior works.}}
\vspace{-5pt}
\footnotesize
\begin{tabular}{lccccc}
\toprule
\makecell[cc]{\textbf{Prior}\\\textbf{Works}}       & \makecell[cc]{\textbf{One}\\\textbf{Mono}} & \makecell[cc]{\textbf{Multi}\\\textbf{Duplicate}} & \makecell[cc]{\textbf{Multi}\\\textbf{Diverse}} & \makecell[cc]{\textbf{Workload}\\\textbf{Assignment}} & \makecell[cc]{\textbf{Specializa}\\\textbf{-tion for Acc}}  \\
\midrule
\makecell[cc]{Eyeriss etc.\\\cite{chen2016eyeriss}-\cite{fccm16}}&\textcolor{ForestGreen}{\checkmark}&\textcolor{red}{$\times$}&\textcolor{red}{$\times$}&\textcolor{red}{$\times$}&\textcolor{red}{$\times$}\\
\arrayrulecolor{black!30}\midrule
\makecell[cc]{DPU etc.\\~\cite{VersalDPU,fpga21mocha}}&\textcolor{ForestGreen}{\checkmark} & \textcolor{ForestGreen}{\checkmark} &\textcolor{red}{$\times$}&\textcolor{red}{$\times$}&\textcolor{red}{$\times$}\\
\midrule
\makecell[cc]{DNN Expl.\\etc.~\cite{zhang2018dnnbuilder,zhang2020dnnexplorer}} & \textcolor{ForestGreen}{\checkmark} & \textcolor{ForestGreen}{\checkmark} & \textcolor{ForestGreen}{\checkmark} &\textcolor{red}{$\times$}&\textcolor{red}{$\times$}\\
\midrule
\makecell[cc]{Herald~\cite{kwon2021heterogeneous}}& \textcolor{ForestGreen}{\checkmark} &  \textcolor{ForestGreen}{\checkmark} & \textcolor{ForestGreen}{\checkmark}&\textcolor{ForestGreen}{\checkmark} &\textcolor{red}{$\times$}\\
\arrayrulecolor{black}\specialrule{0.75pt}{1pt}{3pt}
\makecell[cc]{\textbf{CHARM}\\\textbf{(Ours)}}      &\textcolor{ForestGreen}{\checkmark}&\textcolor{ForestGreen}{\checkmark}&\textcolor{ForestGreen}{\checkmark}&\textcolor{ForestGreen}{\checkmark}&\textcolor{ForestGreen}{\checkmark}\\
\bottomrule
\end{tabular}
    \label{Prior_Comparison}
    \vspace{-5pt}
\end{table}

\section{Versal ACAP Architecture Overview}
\label{sec:arch_overview}

\begin{figure}
\centering
\includegraphics[width=0.8\linewidth]{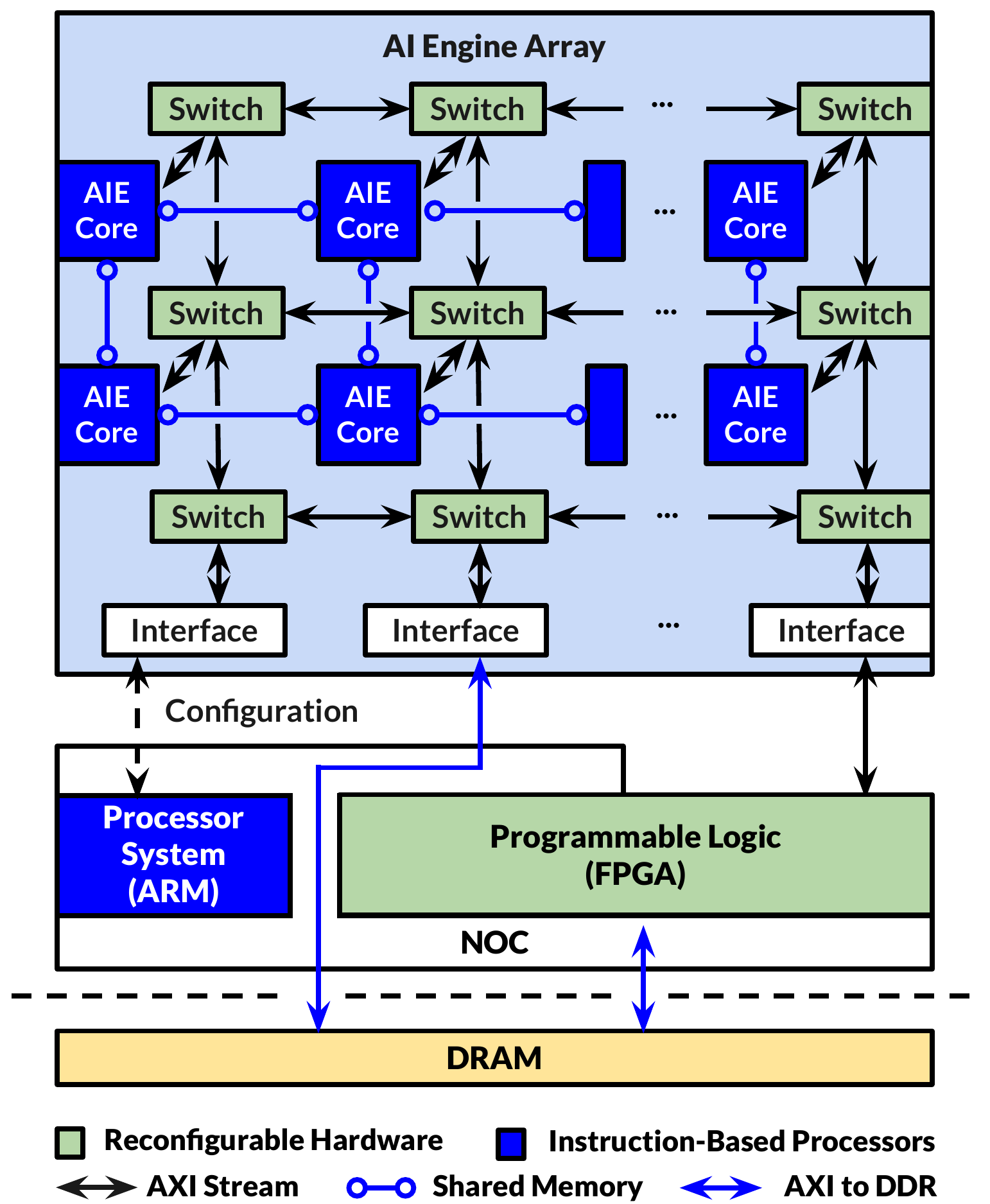}
\caption{Versal ACAP architecture.}
\label{fig:versal-acap-sys-arch}
\vspace{-15pt}
\end{figure}

In this section, we first introduce the system architecture of AMD/\\Xilinx Versal ACAP architecture in Section~\ref{sec:versal_acap_architecture} and then the memory model of AIE array in Section~\ref{sec:aie_comm}. 

\subsection{Versal ACAP Architecture}
\label{sec:versal_acap_architecture}
Figure~\ref{fig:versal-acap-sys-arch} illustrates the overall architecture of the VCK190~\cite{vck190} board and highlights the AIE array on the top.  
The VCK190 board features (1)~the first-generation AIE architecture, which has $8 \times 50$ \textbf{1~GHz} 7-way VLIW processors supporting vector operations up to 1024~bits \cite{aiearch}, (2)~ARM processors to run Linux and general-purpose applications, and (3)~PL to design application-specific hardware with Digital Signal Processors~(DSP) available for integration.  
The AI engine cores and ARM CPUs can be programmed with C/C++ code, while PL can be programmed using both RTL and C/C++ code using High-Level Synthesis~(HLS)~\cite{cong2011high,cong2022fpga,papakonstantinou2009fcuda,papakonstantinou2011multilevel,liang2012high, tapa21}.  
These three components are integrated with I/O peripherals, such as PCIe and DRAM controllers, into a heterogeneous SoC with a Network-on-Chip (NoC). The VCK190 board is equipped with one DDR4-DIMM off-chip memory with a 25.6~GB/s peak bandwidth.

\subsection{AIE Memory Model}
\label{sec:aie_comm}
Each AIE processor tile contains 32~KB of data and is capable of sharing data with the adjacent AIEs in four directions (AIE$\leftrightarrow$neighbor AIE). In addition to local memory shared with adjacent tiles, each AIE tile also connects to an AXI-Stream~(AXIS) switch network, which enables non-local communication between AIE processors (AIE$\leftrightarrow$non-local) and communication with the PL through the PLIOs in the 39 interface tiles (PL$\leftrightarrow$AIEs). The VCK190 board provides 1.2 TB/s (PL$\leftrightarrow$AIEs) / 0.9 TB/s (AIEs$\leftrightarrow$PL)  bandwidth between PL and AIEs, which is ~46$\times$ more than the bandwidth between DDR4 and PL.
The AXIS switches support both circuit-switched and packet-switched connections between ports.  Circuit-switched connections provide dedicated, deterministic communication and support broadcast, where data from a single input channel is transmitted to multiple output channels simultaneously. Packet-switched connections allow data from an input channel to be dynamically routed to different destinations based on a destination header at the start of each packet. This enables data flows to be time-multiplexed on a single routing path. One situation in which we can use packet-switched connections happens when the computation-to-communi\-cation (CTC) ratio of an AIE is more than one. During the computation of AIE 0, the port assigned to this AIE is idle and thus can be used to transfer data to another AIE, say AIE 1, by assigning a different header that matches the destination ID of AIE 1.

\section{CHARM Single Accelerator Design}
\label{sec:mm_design}
In this section, we describe the dataflow and mapping strategy for a single MM acc using hundreds AIEs in Section~\ref{sec:tiling strategy}. Then, in Section~\ref{sec:dataReuse}, we present the data reuse optimizations to balance the massive computation parallelism and communication among AIEs and between PL$\leftrightarrow$AIEs and PL$\leftrightarrow$DDR.

\begin{figure}
\begin{lstlisting}[label=code:Data flow ,caption=Pseudocode of MM loop tiling and dataflow.]
    // Off-Chip <-> On-Chip Time Loop
    for(int i.0=0;i.0<TX;i.0++)   // TX=M/(TI*A*X)
    for(int j.0=0;j.0<TZ;j.0++)     // TZ=N/(TJ*C*Z)
    for(int k.0=0;k.0<TY;k.0++)       // TY=K/(TK*B*Y)
      copyDataFromOffChipOnChip(...)
      // PL On-chip Buffer Reuse Time Loop
      for(int i.1=0;i.1<X;i.1++)   // X
      for(int j.1=0;j.1<Z;j.1++)     // Z
      for(int k.1=0;k.1<Y;k.1++)       // Y
        copyDataFromOnChiptoAIE(...)
        // AIE Array Spatial Loop
        for(int i.2=0;i.2<A;i.2++)   // A
        for(int j.2=0;j.2<C;j.2++)     // C
        for(int k.2=0;k.2<B;k.2++)       // B
            // Single AIE 2D-SIMD Vectorization Loop
            for(int i.3=0;i.3<TI;i.3++)
            for(int j.3=0;j.3<TJ;j.3++)
            for(int k.3=0;k.3<TK;k.3++)
                ...
                2D-SIMD(i.3,j.3,k.3);
\end{lstlisting}
\vspace{-5pt}
\end{figure}

\subsection{Dataflow and Mapping Strategy of a Single Matrix Multiply Accelerator}
\label{sec:tiling strategy}
Listing~\ref{code:Data flow} depicts the overall four-level tiling and mapping strategy for a basic dense matrix-matrix multiply. 
The innermost loop tiling (Lines 16-20) implements MM on a single AIE core and exploits instruction-level parallelism and data-level parallelism by issuing fully pipelined 2D-SIMD (vector-matrix multiplication) instructions. Each AIE stores a (TI$\times$TK) LHS and 
a (TK$\times$TJ) RHS matrix and computes a (TI$\times$TJ) output matrix in its local memory. 
The second-innermost loop tile (Lines 12-14) represents the spatial distribution of execution across different AIE cores in the AIE array. 
These loops are fully unrolled and computed on (A$\times$B$\times$C) AIE cores in a parallel fashion.  
The spatial distribution also corresponds to the number of required IOs, which will be discussed in Section~\ref{sec:dataReuse}. 
The third-innermost time loop tile (Lines 7-9) represents the  sequential processing of data stored in PL on-chip memories. The data from on-chip PL buffers are fed into the AIE array (X$\times$Y$\times$Z) times, and the intermediate partial sum from the AIE array is accumulated on PL.
The outermost loop (Lines 2-4) represents the temporal processing of data stored in off-chip memory, enabling the processing of large matrices that do not fit in on-chip memory. The loop boundary can be decided by the overall input matrix size (M, K, N).

\begin{figure}
     \vspace{-5pt}
     \centering
     \includegraphics[width=0.8\linewidth]{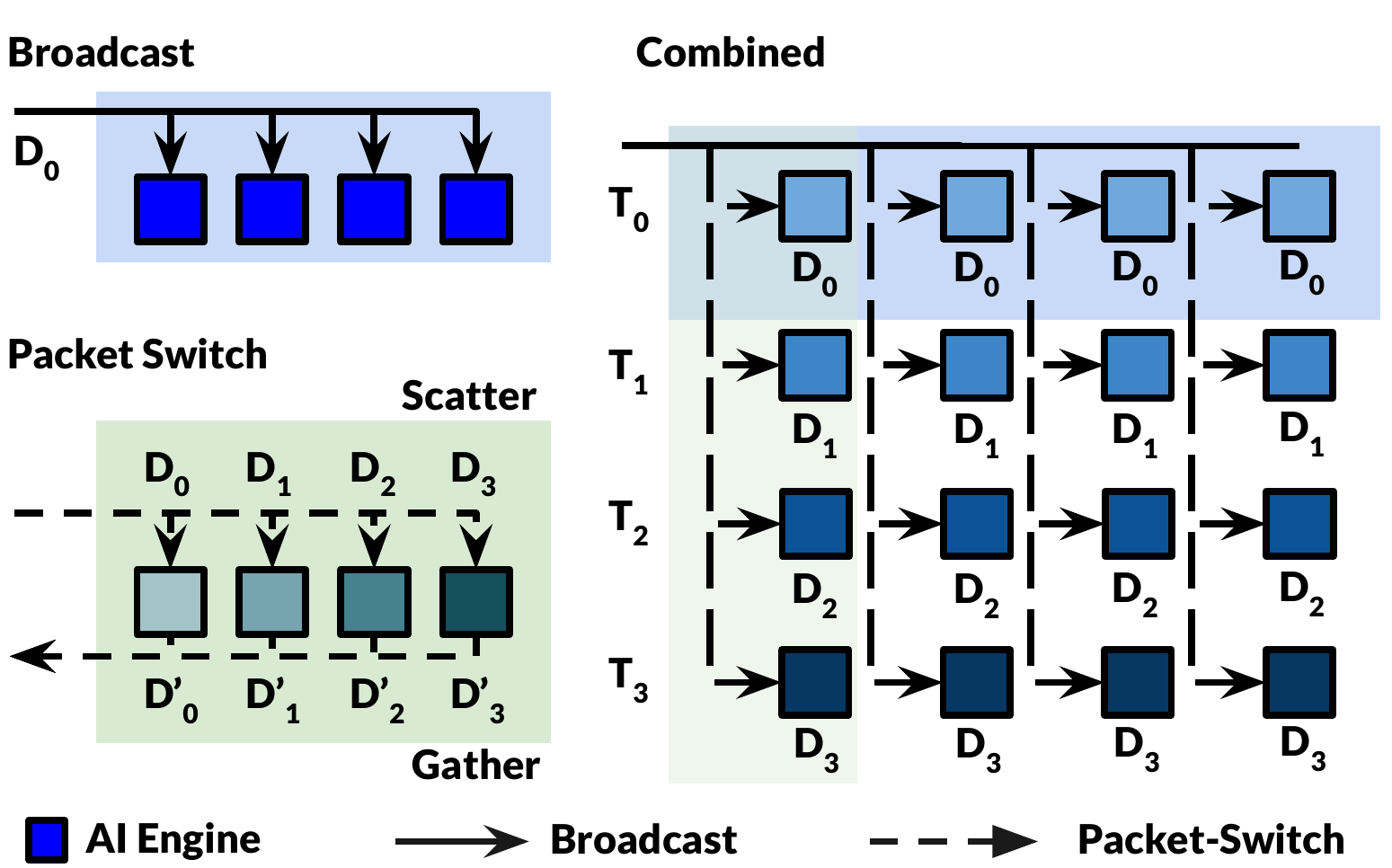}
     \caption{Combining broadcast circuit-switched and packet-switched connections to reduce required I/O to AIE array.}
     \label{fig:AIE IO design}
     \vspace{-10pt}
\end{figure}



\subsection{Data Reuse in Multiple Levels}
\label{sec:dataReuse}
When designing each acc, we adopt a bottom-up strategy and explore data reuse at each level. 
Firstly, at a single AIE level, we make full use of the seven-way VLIW capability of the AIE vector processor to achieve fully pipelined MAC operations by reusing the AIE local register and the local memory.

Secondly, at the PL$\leftrightarrow$AIEs level, when feeding data to tens or hundreds of AIEs, as the number of PLIOs connecting the AIE array and PL is much smaller than the total number of AIE cores, we reduce the number of required PLIO by exploring the data broadcast and packet-switch (described in Section~\ref{sec:aie_comm}) mechanism.
Figure~\ref{fig:AIE IO design} shows how we reuse one single PLIO port by combing broadcast with packet-switch. 
Assume that we have a 4$\times$4 AIE array that calculates an MM with size 1$\times$4$\times$4 (1 MAC/AIE), and it takes one cycle for one AIE to get the left-hand-side (LHS) and the right-hand-side (RHS) operands and four cycles to finish one multiplication which makes the CTC ratio equal to 4. 
By leveraging the data reuse opportunity in MM (e.g., the row of LHS can be reused by different columns of RHS), we can broadcast the first data from LHS to the first row of AIE arrays at Time 0 utilizing one PLIO port as shown in solid lines. 
At Time 1, by specifying  a different destination header, we can transfer the second data of LHS to the second row of the AIE array by reusing the same PLIO port. 
At Time 2 and Time 3, the third and fourth data of LHS are sent to the third and fourth rows of AIEs. 
At Time 4, the first row of AIEs finishes the computation, and the PLIO completes the data transfer to the fourth row of AIEs. 
Therefore, in this case, we can use one PLIO port to send LHS data to 16 AIEs without any performance degradation.

Thirdly, in PL$\leftrightarrow$DDR, 
we further allocate three sets of on-chip buffers for each acc to store the LHS, the RHS, and the output matrices so that a tile of LHS with size (X$\times$A$\times$TI) $\times$ (Y$\times$B$\times$TK) can be reused on-chip for (Z$\times$TJ) times. The buffer size and reuse rate for RHS and output matrices can be calculated in the same way. 
Besides, the double-buffering technique is applied to three buffers to overlap the off-chip data movement with the computation. 
By greatly exploring the data reuse opportunities at multiple levels, our system can sustain high computation efficiency under limited off-chip bandwidth, i.e., 25.6 GB/s of DIMM-DDR4 on VCK190.


\section{CHARM Architecture and CHARM Framework to Compose Multiple Diverse Accelerators}
\label{sec:dse_new}

\begin{figure}
    \centering
    \includegraphics[width=1\linewidth]{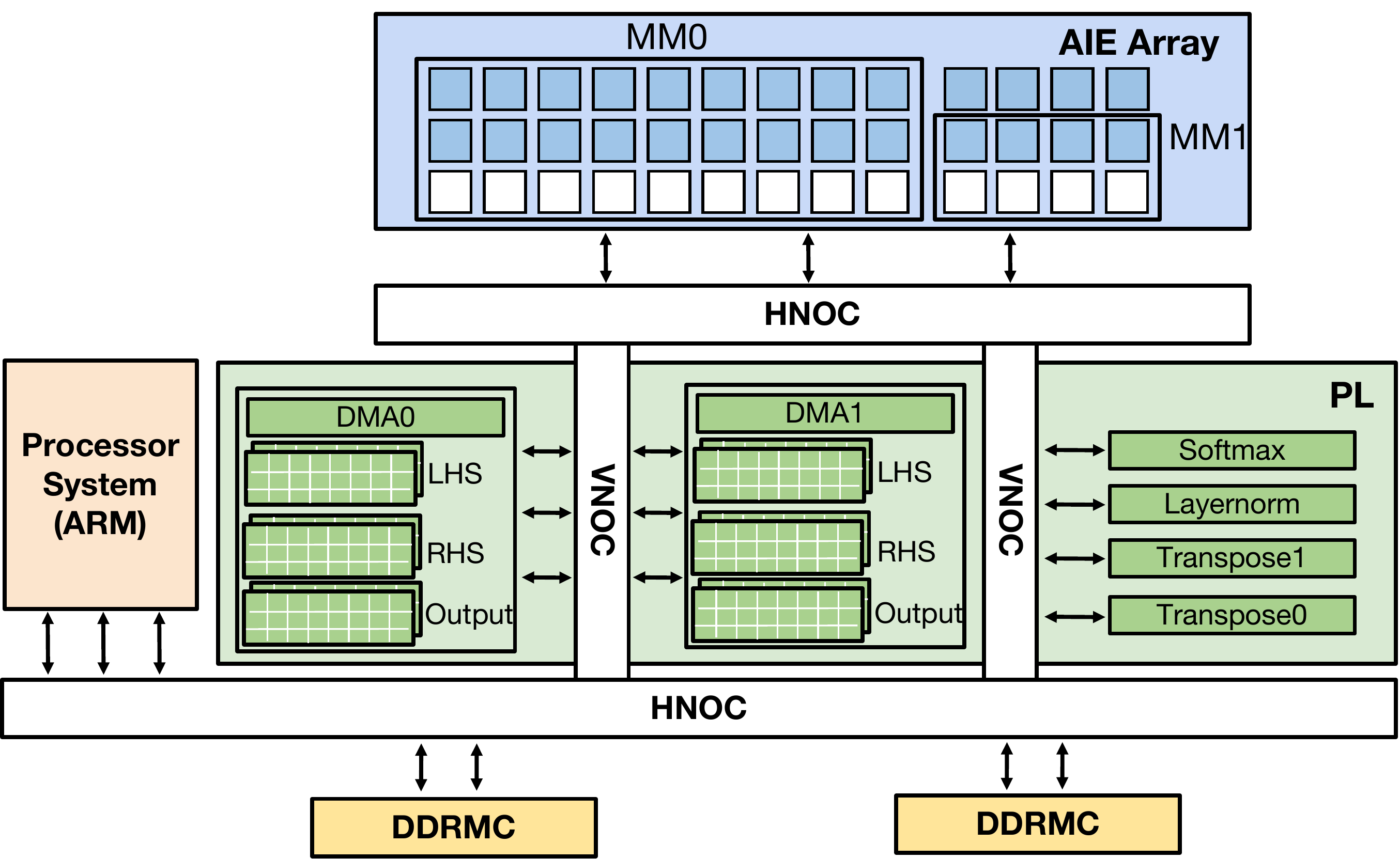}
    \vspace{-10pt}
    \caption{System architecture of multiple diverse MM accs and other non-MM accs.}
    \label{System_design}
    \vspace{-5pt}
\end{figure}

In this section, we introduce the CHARM architecture in Section~\ref{sec:Proposed_arch} and CHARM framework overview in Section~\ref{sec:charm_frame_overview}. 
We then discuss each module within the framework from Section~\ref{sec:charm_DSE} to Section~\ref{sec:charm_RTS}.

\subsection{CHARM Architecture}
\label{sec:Proposed_arch}

Figure ~\ref{System_design} illustrates the CHARM architecture with one or more diverse MM accs in the system and other kernel accs for non-MM kernels within an end-to-end deep learning application.
We partition the AIE array for multiple MM accs (two in this example). 
For each MM acc, we design a specialized DMA module that contains the data transferring control logic and on-chip buffer according to the tiling strategy. 
The different AIE partitions communicate with their corresponding DMA modules through the PLIO interface and NOC. 
We refer to the AIE array, its corresponding PLIO, and the DMA module as one MM acc design. 
For each non-MM kernel, e.g., transpose, softmax, and layer normalization in BERT, ViT models, we design one acc for each type of kernel on the PL side.
Each non-MM acc contains DMA, computation logic, and local buffers.
For these communication-bound kernels, the design goal is to achieve near-peak off-chip bandwidth.
When running these kernels, as they consume all the off-chip bandwidth, we choose to sequentially launch these non-MM and communication-bound kernels before or after MM acc(s).


\subsection{CHARM Framework Overview}
\label{sec:charm_frame_overview}
\begin{figure}
    \vspace{-5pt}
    \centering
    \includegraphics[width=1\linewidth]{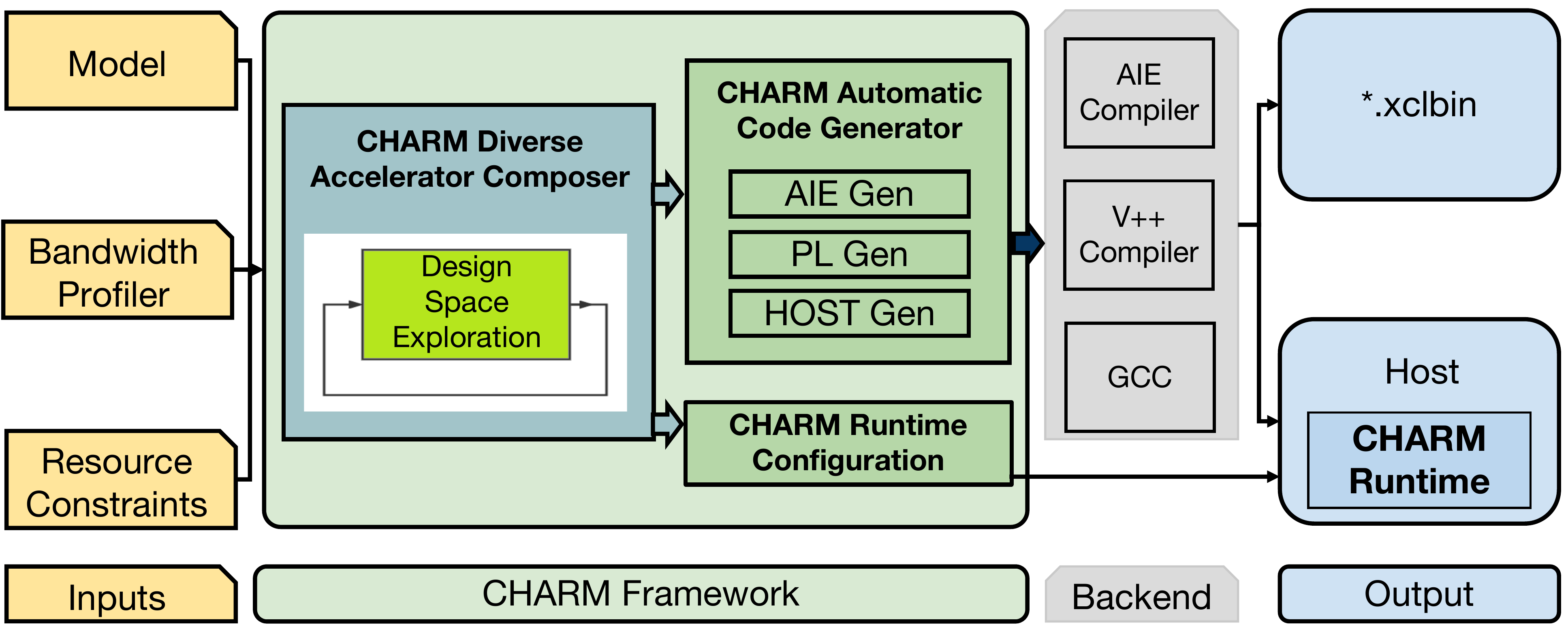}
    \vspace{-10pt}
    \caption{CHARM framework overview.}
    \label{fig:CHARM_Framework}
    \vspace{-10pt}
\end{figure}

We illustrate the proposed CHARM framework overview in Figure ~\ref{fig:CHARM_Framework}. 
The CHARM framework takes the application model, platform off-chip bandwidth profiling, and platform hardware resource constraints as input, performs automated optimization and code generation, and launches backend compilers to generate the ready-to-run binaries as output. 
There are several modules in the CHARM framework: (1) On top of CDSE, CDAC finds the optimal design with the highest throughput and outputs the configurable design parameters for each acc. It also generates a runtime config file that specifies which acc should be called for a certain kernel. (2) CACG takes the  configurable parameters generated from CDAC as input, implements the design, and generates all the needed source code files for AIEs, PL, and host CPUs. 
CHARM calls the corresponding backend tools to generate both the hardware bitstream and host binaries.
(3) CRTS takes the runtime config files from CDAC and kernel dependency graph as input and schedules the kernels in the task pools onto available accs.


\subsection{CHARM Design Space Exploration (CDSE) for a Single Acc}
\label{sec:charm_DSE}
\noindent\textbf{DSE Configurable parameters: A, B, C, X, Y, Z.} 
In order to attain optimized throughput for each diverse accelerator, we design CDSE, which takes matrix sizes (M, K, and N), optional user-specified hardware constraints, and hardware platform off-chip characterization database as input and perform an analytical model-based search. 
During CDSE, we set the single AIE workloads to 32$\times$32$\times$32, i.e., TI=TK=TJ=32.  We achieve up to 95\% of kernel efficiency for MM, utilize 75\% of the AIE local memory in this design point, and obtain the CTC ratio of 4. 
The outputs of CDSE are the configurable parameters, including \textbf{A, B, C, X, Y, Z} that meet all the hardware constraints. 
The parameters {A, B, C} determine the number of AIE and PLIO used in the AIE array. 
{X, Y, Z, A, B, C} together with pre-fixed parameters \textbf{TX, TY, TZ} decide the number of utilized on-chip buffers. 
This optimization problem can be formulated as an integer programming~(IP) optimization problem shown as below. $\mathit{AIE_{num}}$, $\mathit{PLIO_{in}}$, $\mathit{PLIO_{out}}$ and $\mathit{On\_chip_{RAM}}$ represent the user-specified hardware constraints:
\begin{equation}
\label{eq:Throughput}
\max   \mathit{Throughput} = \mathit{M} \cdot \mathit{K} \cdot \mathit{N} \cdot 2 /  \mathit{TIME}\\
\end{equation}
\begin{equation}
\label{eq:AIE Tiling}
\textrm{s.t.}  A\times B\times C \le \mathit{AIE_{num}}\\
\end{equation}

\vspace{-10pt}
\begin{eqnarray}
\small{
\begin{gathered}
\label{eq:port in}
\mathit{Port_{in}} \le \mathit{PLIO_{in}},\\   
\mathit{Port_{out}} \le \mathit{PLIO_{out}}\\
\end{gathered}
}
\end{eqnarray}

\vspace{-0.1in}
\begin{equation}
\small{
\mathit{Buff}  \le \mathit{On\_chip_{RAM}}
}
\end{equation}

\noindent\textbf{AIE-Array Tiling Selection.}
Since $\{$A, B, C$\}$ are fully unrolled and mapped to the AIE Array, the multiplication of the unroll factors A, B, and C should be less than or equal to the total number of AIEs in Equation~\ref{eq:AIE Tiling}. 
The number of packet-switch ports is determined by $\{$A, B, C$\}$ and the I/O reuse mechanism described in Section~\ref{sec:dataReuse}. 
They should meet the input and output PLIO resource constraints. The input and output PLIO numbers can be obtained by: 
\begin{eqnarray}
\small{
\begin{gathered}
\label{eq:PLIO}
    \mathit{Port}_{in} =  \lceil A \cdot  B/\mathit{CTC}\rceil \\+  \lceil C \cdot  B/\mathit{CTC}\rceil\\
    \mathit{Port}_{out} = \lceil  A \cdot C /\mathit{CTC}\rceil
    \vspace{-0.02in}
\end{gathered}
}
\end{eqnarray}


\noindent\textbf{PL Tiling Selection.} On-chip PL buffers are allocated in order to amortize the 46x bandwidth gap between off-chip to PL and PL to AIE-Array by increasing the data reuse rate. 
Equation \ref{eq:buffer} shows the size of LHS, RHS, output buffers, and their off-chip to on-chip communication time. BPD refers to bytes per data and BW\_{L,R,O} are the off-chip bandwidth measured from bandwidth profiling.

\vspace{-0.1in}
\begin{eqnarray}
\small{
\begin{gathered}
\label{eq:buffer}
\mathit{Buff_{L}} =( X \cdot A \cdot TI) \cdot (Y \cdot \mathit{B \cdot TK}) \cdot \mathit{BPD} \\
\mathit{Buff_{R}} = Y \cdot Z \cdot B \cdot C \cdot \mathit{TK \cdot TJ} \cdot \mathit{BPD} \\
\mathit{Buff_{O}} = X \cdot Z \cdot A \cdot C \cdot \mathit{TI \cdot TJ} \cdot \mathit{BPD} \\
\mathit{Buff}= 2 \cdot (\mathit{Buff}_{L} + \mathit{Buff}_{R} + \mathit{Buff}_{O}) \\
\mathit{Time_{L,R,O}} = {\mathit{Buff}_{L,R,O}} / {\mathit{BW}_{L,R,O}} \\
\end{gathered}
}
\vspace{-0.25in}
\end{eqnarray}

\noindent\textbf{Performance Modeling.} To calculate the overall execution time, the scheduling of data communication between off-chip to on-chip and the AIE array computation should be considered. 
The computation time for all the on-chip time loops, i.e., Line 6 in Listing~\ref{code:Data flow}, can be defined by Equation ~\ref{eq:compute} in which MAC represents the theatrical MAC operation that one AIE engine can do in one cycle, and Eff refers to the real efficiency that the computation kernel achieves. 
We consider both single AIE and AIE array pipeline efficiency (PL$\leftrightarrow$AIE) here and assign the overall efficiency to 80\%. 
For the off-chip to on-chip scheduling, as described in Listing ~\ref{code:Data flow}, the loop order of the outermost loop is TY$\rightarrow$TZ$\rightarrow$TX, thus, the memory access time for LHS and RHS will happen TX$\times$TX$\times$TZ times in total.
The overall execution TIME can be calculated by Equation ~\ref{eq:overall_time}. 
This is an equation for illustration purposes where we leave out the details on the formulation of time spent storing the output and prologue and epilogue time in the pipeline.
\begin{equation}
\small{
\label{eq:compute}
\mathit{Time\_comp} = (X \cdot Y \cdot Z \cdot TI \cdot TK \cdot TJ / \mathit{MAC}) / \mathit{Eff}
}
\vspace{-10pt}
\end{equation}

\vspace{-0.05in}
\begin{eqnarray}
\small{
\begin{gathered}
\label{eq:overall_time}
    \mathit{TIME} =  max([\mathit{Time_{L}},\mathit{Time_{R}},\mathit{Time\_comp}]) \\\cdot (TX \cdot TY \cdot TZ) \\
\end{gathered}
}
\vspace{-0.25in}
\end{eqnarray}

For any specific shape(s), all the possible configurable parameters will be evaluated in an exhaustive fashion. 
After CDSE, top-ranked optimized design points will be reported. 

\subsection{CHARM Diverse Acc. Composer (CDAC)}
\label{sec:charm_DAC}

\noindent\textbf{Two-step search algorithm in CDAC.} 
To achieve overall optimized throughput when mapping diverse sizes of MM kernels on multiple accs, we propose a sort-based two-step algorithm in CDAC. 
In the first step of CDAC, we partition the MM kernels of different workloads within an input model to multiple groups. The number of groups equals the number of diverse accs, which is a hyperparameter in CDAC.
After the workload partition, in the second step, we generate a resource partition candidate that specifies the resource budget for each accelerator to be proportional to the total number of operations from the assigned MM kernel(s). 
Under the assigned workload and assigned resource, we search all valid candidates of configurable parameters (A,B,C,X,Y,Z) for each accelerator. We then fine-tune the memory resource partition to generate more resource partition candidates.
After the memory fine-tuning, we generate a new workload partition and redo the resource partition and configurable parameter search which further optimize the system throughput of all the accs. We discuss the details of each step as follows.

\begin{algorithm}
\caption{Diverse Accelerator Composer Algorithm}
\label{alg:Acc_composer}
\footnotesize
\begin{flushleft}
\textbf{Input: layer[n], bw, hw\_sr, num, ubound} \\

\Comment{\textcolor{magenta}{\textbf{layer[n]} are n layers in an application. \textbf{bw} refers to bandwidth, \textbf{hw\_sr} includes the AIE, PLIO, RAM resources, \textbf{num} refers to the number of accs, \textbf{ubound} is the hyperparameter for memory tuning}}\\
\textbf{Output: Workload[num], final\_Acc[num]}  \\
\Comment{\textcolor{magenta}{\textbf{Workload} and \textbf{final\_Acc} contains the workload assignment and the hardware configuration for each acc respectively}}
\end{flushleft}

\begin{algorithmic}[1]
\State $\mathit{BW} \gets \mathit{bw}/\mathit{num\_acc}$
\State $\mathit{HW.RAM}[:] \gets   \mathit{hw\_sr.ram}/\mathit{num\_acc}$
\State $\mathit{final\_cycle} \gets \mathit{inf}$
\State $\mathit{layer\_sort}[:] \gets \mathrm{sort}(\mathit{layer})$
\For{sche in range($C\mathit{\binom{n-1}{\mathit{num}-1}}$)}
    \State $\mathit{partition}[:] \gets \mathrm{partition}(\mathit{layer\_sort}[:],\mathit{num},\mathit{sche})$ \Comment{\textcolor{magenta}{1st step}}
    \State $\mathit{op\_portion}[:] \gets \mathrm{cnt}(\mathit{partition}[:])$
    \Comment{\textcolor{magenta}{2nd step}} 
    \State $\mathrm{update}(\mathit{HW.AIE}[:],\mathit{HW.PLIO}[:],\mathit{op\_portion}[:])$
    \State $\mathit{Acc}[:], \mathit{cycle}[:] \gets \mathrm{Acc\_search}(\mathit{HW},\mathit{BW},\mathit{partition}[:])$ \\
    \Comment{\textcolor{magenta}{Sequentially launch CDSE}}
    \While{$ \mathit{tune\_cnt} \neq \mathit{ubound}$} \Comment{\textcolor{magenta}{Memory tuning}}
        \State $\mathit{index} \gets \max(\mathit{cycle}[:])$ 
        \State $\mathrm{update}(\mathit{HW.RAM}[:],\mathit{index})$ \Comment{\textcolor{magenta}{Increase the memory of the slowest acc}}
        \State $\mathit{Acc}[:], \mathit{cycle}[:] \gets \mathrm{Acc\_search}(\mathit{HW},\mathit{BW},\mathit{partition}[:])$
        \If{$\max(\mathit{cycle}[:]) < \mathit{final\_cycle}$ }\Comment{\textcolor{magenta}{Update optimal point}}
            \State $\mathit{final\_cycle} \gets \max(\mathit{cycle}[:]) $
            \State $\mathit{final\_Acc}[:] \gets \mathit{Acc}[:]) $
            \State $\mathit{Workload}[:] \gets \mathit{partition}[:] $
        \EndIf
        \State $\mathit{tune\_cnt}\mathit{++}$
    \EndWhile
    
\EndFor

\State Define $\mathrm{Acc\_search}(\mathit{HW},\mathit{BW},\mathit{partition}[:]):$
    \For{acc in range(num)}
        \State $\mathrm{CDSE}(\mathit{partition}[\mathit{acc}],\mathit{HW}[\mathit{acc}],\mathit{BW}[\mathit{acc}])$
    \EndFor
    \State $\mathrm{return}~\mathit{Acc}[:], \mathit{Cycle}[;]$
\end{algorithmic}
\end{algorithm}


\noindent\textbf{1st Step: Workload Assignment}. To improve the overall throughput of the diverse acc architecture, we need to properly assign the MM kernels to the accs and make them work concurrently with a similar execution time. 
However, mapping an application with \textbf{n} kernels to \textbf{num} accs suffers from the exponential time complexity as the total mapping search space scales as $\mathcal{O}(num^{n})$ 
. 
To better scale larger models that contain more kernels, i.e., a larger n, we propose a sort-based algorithm to partition the workload with reduced time complexity as $\mathcal{O}(\binom{n-1}{num-1})=C^{n-1}_{num-1}$. 
As shown in Algorithm~\ref{alg:Acc_composer}, CDAC first sorts the different shapes of the
MM kernels by their number of operations (Line 4) so that MMs with larger and smaller sizes can be properly divided. 
Then we divide the sorted MM kernels into n groups (Lines 5-6). 
For example, if there are eight different shapes of kernels that need to be mapped to num=2 accs, after sorting the kernels, we put one separator between any two kernels to separate all kernels into two groups. 
In total, it gives us $C\binom{8-1}{2-1}$ = 7 grouping design choices.

\noindent\textbf{2nd Step: Hardware Resource Partitioning}. 
For each workload assignment, we perform DSE to find the optimized acc configurable parameters under the partitioned hardware resource constraints, including the number of AIEs, PLIO, on-chip RAM, and off-chip bandwidth. 
To minimize the maximum execution time of all the accs, CDAC assigns the number of AIEs and PLIO constraints proportional to the total number of operations assigned to the acc (Lines 7-8). 
For the number of on-chip RAMs, we first evenly distribute it (Line 2). 
After sequentially launching CDSE to find the configuration of every acc once (Lines 9-10), we apply a memory fine-tuning step to optimize the memory allocation. 
It finds the index of the acc that consumes the most time (Line 12) and then tries to explore a better configuration by increasing the memory allocation of this acc while decreasing the memory allocation of others'(Line 13-14).
If a better result is found, we update the global optimal execution cycles and corresponding acc configuration settings (Line 15-18). 
Note that, in the current model, we assume each acc evenly occupies the off-chip bandwidth (Line 1) and leave the discussion of the off-chip bandwidth partition for future work.

\vspace{-5pt}
\subsection{CHARM Auto. Code Generation (CACG)}
\label{sec:charm_ACG}
After finding the hardware design parameters of optimized designs from CDAC, we implement CACG, including AIEGen, PLGen, and HostGen, to generate the corresponding source code for AIEs, PL, and host CPU.
AIEGen takes the tiling factor of a single AIE (TI,TK,TJ) and AIE Array (A,B,C) as input and instantiates the corresponding number of AIE cores. 
It leverages the C++-based Adaptive Data Flow (ADF) Graph API~\cite{ADF} to build connections among AIE cores through the AXI network and connections between AIE Array and PL through PLIOs.
Using the PL level (X,Y,Z) design parameters, PLGen generates HLS C/C++ code that allocates on-chip buffers on the PL side and implements the data transferring modules for sending/receiving data to/from the AIE array. 
HostGen emits the Xilinx runtime library (XRT) API-based host code. 

After code generation, CHARM launches the vendor tools, including the AIE compiler and the V++ compiler to generate the output object files \texttt{libadf.a} and \texttt{kernel.xo} which are linked into one \texttt{xclbin}, i.e., the hardware bitstream of the design.
The GCC compiler compiles XRT-API-based host code to host program runs on the ARM CPU for kernel scheduling and system controls.

\vspace{-5pt}
\subsection{CHARM Runtime Scheduler (CRTS)}
\label{sec:charm_RTS}
\vspace{-10pt}

\begin{algorithm}
\caption{Runtime Scheduler Algorithm}
\label{alg:runtime_scheduler}
\footnotesize
\begin{flushleft}
\textbf{Input: Graph, num, task\_pool[task][layer]} \\  
\textbf{Output:} Runtime scheduling for each accelerator  
\end{flushleft}
\begin{algorithmic}[1]
\While{(1)}  \Comment{\textcolor{magenta}{Assign ready tasks to corresponding Accs}}
    \For{$\mathit{acc}$ in range($\mathit{num}$)}
        \If{$\lnot Acc[acc].\mathrm{idle}()$}
            \State \textbf{Continue}
        \EndIf
        \For{$t$ in range($\mathit{tasks}$)}
            \For{$l$ in range($\mathit{layer}$)}
                \If{$\mathit{task\_pool}[t][l] \neq \emptyset
                \land \mathit{task\_pool}[t][l].\mathrm{valid}()$}
                    \State $\mathit{Acc}[\mathit{acc}].\mathrm{assign}(\mathit{task\_pool}[t][l])$
                    \State \textbf{Continue} line 2
                \EndIf
            \EndFor
        \EndFor
    \EndFor
\EndWhile
\While{(1)}  \Comment{\textcolor{magenta}{Update task\_pool according to dependency graph}}
    \For{$\mathit{acc}$ in range($\mathit{num}$)}
        \If{$\mathit{Acc}[\mathit{acc}].\mathrm{finish}()$}
            \State $\mathit{task\_pool}.\mathrm{update}(\mathit{Graph})$
            \State $\mathit{Acc}[\mathit{acc}].\mathrm{update}(\mathit{idle})$
        \EndIf
    \EndFor
\EndWhile
\end{algorithmic}
\end{algorithm}
\vspace{-10pt}

To achieve high throughput while maintaining relatively low latency under dependency constraints within each task, we propose CRTS that runs on the ARM CPU during runtime.
Algorithm~\ref{alg:runtime_scheduler} lists the scheduler algorithm. 
It takes the dependency graph, number of accelerators, and layer assignment configuration file generated by CDAC as input. 
There are two parallel processes in CRTS. 

The first process keeps tracking to check if there are any idle accs we can assign tasks to~(Lines 2-3). 
CRTS traverses the layers assigned to this acc following a first-in-first-out principle~(Lines 5-6). 
If the layer is still in the task pool, it means that it has not been issued. Suppose all the preceding layer(s) of the current layer have been executed, i.e., dependency resolved. In that case, CRTS assigns this valid layer to the corresponding acc~(Lines 7-8) and continues to track other accs~(Line 9).
The second process keeps track of the status of every acc to see if it has finished the workload~(Lines 12-13) and updates the task pool according to the dependency graph, as well as changing the status of the acc~(Line 14) to idle. 

\section{Experiment Results}
\label{sec:experiments}
In this section, we first illustrate the single AIE efficiency and single MM acc throughput in Section~\ref{sec:exp_aie} and ~\ref{sec:exp_one_acc}. 
In Section~\ref{sec:exp_app}, we implement different CHARM designs, including one monolithic MM acc, one specialized MM acc, two-diverse MM accs, and eight-duplicate MM accs for four applications: BERT, ViT, NCF, and MLP. 
All the experiments are conducted on VCK190 with 230MHz on PL and 1GHz on AIE. 
AMD/Xilinx Vitis version 2021.1 is used as the compilation backend tool. When measuring the power consumption, we iterate each application for more than 60s and report the average value by employing the board evaluation and management tool, AMD/Xilinx BEAM~\cite{BEAM}.

\subsection{Single AIE Kernel Efficiency Comparison}
\label{sec:exp_aie}
In this section, we showcase our single AIE MM computation efficiency under different matrix sizes for fp32. 
We leverage the AIE intrinsics ~\cite{AIE_Intrinsics} to program the single kernel design and obtain the execution cycle of our single AIE design by simulating on the Versal ACAP AI Engine System C simulator~\cite{AIE_simulator}, a cycle-accurate architecture simulator.
As shown in Table \ref{tbl:XDSE Acc}, our single AIE can achieve up to 7.57 MACs/cycle and 94.70\% peak performance when MM size equals 32$\times$32$\times$32. 
Compared to the AIE dense MM kernel efficiency reported in H-GCN~\cite{zhang2022h}, our single kernel obtains 2.26$\times$ average efficiency gain.
For the whole system design, we choose 32$\times$32$\times$32 as our single kernel as it achieves high computation efficiency and the total size of LHS, RHS and output matrices are within 16 KB so that they fit in the AIE local memory and can be double buffered.

\begin{table}
\footnotesize
\caption{Single AIE MM comparison under fp32 data type.}
\vspace{-10pt}
\label{tbl:XDSE Acc}
    \begin{center}
\begin{tabular}{ c | c c | c c c }
 \toprule
& \multicolumn{2}{c}{\textbf{\textcolor{black}{H-GCN\cite{zhang2022h}}}}
& \multicolumn{3}{c}{\textbf{\textcolor{black}{CHARM (this work)}}} \\
   \textbf{\textcolor{black}{Size: M x K x N}}  
  &  \textbf{\textcolor{black}{MACs/Cyc}} &  \textbf{\textcolor{black}{Eff}}  & \textbf{\textcolor{black}{MACs/Cyc}} & \textbf{\textcolor{black}{Eff}}
   & \textbf{\textcolor{black}{Eff gain}}\\
    \midrule
\textcolor{black}{ $16\times 16 \times 16$}  &  \textcolor{black}{2.34} &   \textcolor{black}{29.30\%}  &   \textcolor{black}{ 6.18 } &   \textcolor{black}{77.22\%} &   \textbf{\emph{2.64x}}   \\
\textcolor{black}{ $32\times 32 \times 32 $}   &   \textcolor{black}{3.64}  &   \textcolor{black}{45.50\%}   &   \textcolor{black}{ 7.57 } &   \textcolor{black}{94.70\%}  & \textbf{\emph{2.08x}}  \\
\textcolor{black}{ $ 64\times 64 \times 8 $}   &  \textcolor{black}{3.64} &   \textcolor{black}{45.50\%}     &   \textcolor{black}{ 7.54 } &   \textcolor{black}{94.29\%} & \textbf{\emph{2.07x}}   \\

    \bottomrule
\end{tabular}
    \end{center}
    \vspace{-10pt}
\end{table}

\subsection{Performance for Square MMs on One Monolithic Accelerator}
\label{sec:exp_one_acc}

\begin{table}[!tb]
\caption{Performance comparison in GFLOPS between on-board measurements and CDSE analytical modeling estimations  under different matrix sizes. The error rates in percentage show that CDSE achieves a high prediction accuracy.}
\vspace{-10pt}
\label{tbl:CDSE_Accuracy}
\begin{adjustbox}{width=1\columnwidth,center}
\begin{tabular}{c | c c c c c}
\toprule
\textbf{\textcolor{black}Square MM size} & \textbf{\textcolor{black}On-board} & \textbf{\textcolor{black}Estimation} & \textbf{\textcolor{black}Error} & \textbf{\textcolor{black}Power(W)} \\
\midrule
64   & 0.41    & 0.40    & -2\% & 32.58 \\
128  & 3.36    & 3.22    & -4\% & 32.86 \\
256  & 25.58   & 25.79   & 1\%  & 34.66 \\
512  & 176.24  & 178.42  & 1\%  & 37.95 \\
1024 & 1103.46 & 1123.81 & 2\%  & 41.78 \\
1536 & 1633.13 & 1649.01 & 1\%  & 46.02 \\
2048 & 1672.76 & 1688.17 & 1\%  & 47.87 \\
3072 & 2850.13 & 2895.90 & 2\%  & 50.65 \\
4096 & 2718.42 & 2773.26 & 2\%  & 51.97 \\
6144 & 3277.99 & 3363.89 & 3\%  & 53.57 \\
\bottomrule
\end{tabular}
\end{adjustbox}
\end{table}

We evaluate the throughput of one monolithic acc design and compare the performance between the modeling estimation from CDSE and the on-board measurement. 
We build the monolithic design by using 384 AIEs and over 83\% of on-chip RAM utilization with the AIE running at 1GHz and the PL side at 230MHz. 
As shown in Table~\ref{tbl:CDSE_Accuracy}, the throughput of the one-acc monolithic design rises as the square MM size increases. 
While it achieves 3.27 TFLOP/s at size 6144, the throughput at size 64 is only 0.41 GFLOPs. 
CHARM CDSE is capable of precisely estimating the on-board execution time with an average estimation error rate of only 2.9\%.

We compare the throughput of the same MM application implemented only in the PL part of VCK190 using the state-of-the-art systolic-array-based framework AutoSA~\cite{Wang_Guo_Cong_2021} for fp32 data type. 
The PL side of VCK190 is featured with 1968 DSP58 IPs.
Instead of using five DSP48 to calculate the floating point multiplication in the previous generation board, it only consumes one DSP58. 

As shown in Table~\ref{tbl:comparison_vck190_autosa}, the CHARM single MM acc achieves 3.27 TFLOPs throughput, 5.54$\times$ throughput and 1.93$\times$ energy efficiency gains compared to the PL-only design on VCK190.

\begin{table}
\footnotesize
\caption{\textcolor{black}{Comparison between PL only design and PL RAM + AIE design in CHARM on VCK190.}}
\vspace{-10pt}
\label{tbl:comparison_vck190_autosa}
\begin{adjustbox}{width=0.6\columnwidth,center}
\begin{tabular}{ c | c  c }
 \toprule
& \multicolumn{1}{c}{\textbf{\textcolor{black}{PL~\cite{Wang_Guo_Cong_2021}}}}
& \multicolumn{1}{c}{\textbf{\textcolor{black}{CHARM}}} \\
   \textbf{\textcolor{black}{Data Type}}  
  &  \textbf{\textcolor{black}{Float32}} & \textbf{\textcolor{black}{Float32}}
\\
    \midrule
\textcolor{black}{ Frequency}    &   \textcolor{black}{ PL:200MHz}    &  \textcolor{black}{PL:230MHz}   \\
\textcolor{black}{ URAM }        &   \textcolor{black}{ 0 }           &  \textcolor{black}{384}    \\
\textcolor{black}{ BRAM }        &   \textcolor{black}{ 923 }           &  \textcolor{black}{764}    \\
\textcolor{black}{ DSP/AIE }     &   \textcolor{black}{ DSP58:1536}   &  \textcolor{black}{AIE:384}    \\
\textcolor{black}{ TFLOPs}         &   \textcolor{black}{ 0.59 (1x)}    &  \textcolor{black}{3.27 (5.54x)} \\
\textcolor{black}{ Power(W)}     &   \textcolor{black}{ 18.60 }       &  \textcolor{black}{53.40}       \\
\textcolor{black}{ Energy Eff}   &   \textcolor{black}{ 1x}           &  \textcolor{black}{1.93x}        \\
    \bottomrule
\end{tabular}
\vspace{-10pt}
\end{adjustbox}
\end{table}

\subsection{End-to-End Performance}
\label{sec:exp_app}

\begin{table}[!tb]
\caption{MM sizes in BERT, ViT, NCF, MLP.}
\vspace{-10pt}
\label{tbl:Evaluation}
\begin{center}
\begin{adjustbox}{width=0.8\columnwidth,center}
\begin{tabular}{c | c c c c c}
\toprule
\textbf{\textcolor{black}Model} & \textbf{\textcolor{black}\# of layer} & \textbf{\textcolor{black}M} & \textbf{\textcolor{black}K} & \textbf{\textcolor{black}N} & \textbf{\textcolor{black} batch dot size} \\
\midrule
\multirow{5}{*}{BERT}
& 4 & 3072  & 1024  & 1024 & N/A\\ 
& 1 & 3072  & 4096  & 1024 & N/A\\
& 1 & 3072  & 1024  & 4096 & N/A\\
& 1 & 512  & 64  & 512 & 96\\
& 1 & 512  & 512  & 64 & 96\\
\midrule
\multirow{6}{*}{ViT} 
& 1 & 3072  & 3024 & 1024 & N/A\\ 
& 1 & 3072  & 1024 & 1024 & N/A\\ 
& 1 & 3072  & 1024 & 4096 & N/A\\ 
& 1 & 3072  & 4096 & 1024 & N/A\\ 
& 1 & 3072  & 1024 & 3048 & N/A\\ 
& 2 & 64  & 64 & 64 & 768\\ 
\midrule
\multirow{9}{*}{NCF} 
& 1 & 3072  & 4096 & 2048 & N/A\\ 
& 1 & 3072  & 2048 & 1024 & N/A\\ 
& 1 & 3072  & 1024 & 512 & N/A\\ 
& 1 & 3072  & 512 & 256 & N/A\\ 
& 1 & 3072  & 256 & 128 & N/A\\ 
& 1 & 3072  & 128 & 64 & N/A\\ 
& 1 & 3072  & 64 & 32 & N/A\\ 
& 1 & 3072  & 32 & 16 & N/A\\ 
& 1 & 3072  & 32 & 1 & N/A\\
\midrule
\multirow{3}{*}{MLP} 
& 1 & 3072  & 2048  & 4096 & N/A\\ 
& 2 & 3072  & 4096  & 4096 & N/A\\ 
& 1 & 3072  & 4096  & 1024 & N/A\\ 
\bottomrule
\end{tabular}
\end{adjustbox}
\end{center}
\end{table}


\begin{table}[!tb]
\scriptsize
\caption{Time breakdown for different types of kernels in the end-to-end solutions that achieves the highest throughput for BERT, ViT, NCF and MLP.}
\vspace{-10pt}
\label{tbl:time_breakdown}
\begin{adjustbox}{width=0.8\columnwidth,center}
\begin{tabular}{c | c c c c}
\toprule
\textbf{\textcolor{black}Kernel} & \textbf{\textcolor{black}BERT} & \textbf{\textcolor{black}ViT} &
\textbf{\textcolor{black}NCF} &
\textbf{\textcolor{black}MLP} \\
\midrule
MM        & 57.2ms & 57.7ms & 40.4ms & 11.9ms    \\
Layernorm & 4.5ms  & 4.5ms  & 0 & 0   \\
Softmax   & 18.7ms & 2.3ms  & 0  & 0  \\
Transpose & 5.2ms  & 5.2ms  & 0  & 0 \\
\bottomrule
\end{tabular}
\end{adjustbox}
\vspace{-10pt}
\end{table}

\begin{figure}
    \centering
    \includegraphics[width=0.7\linewidth]{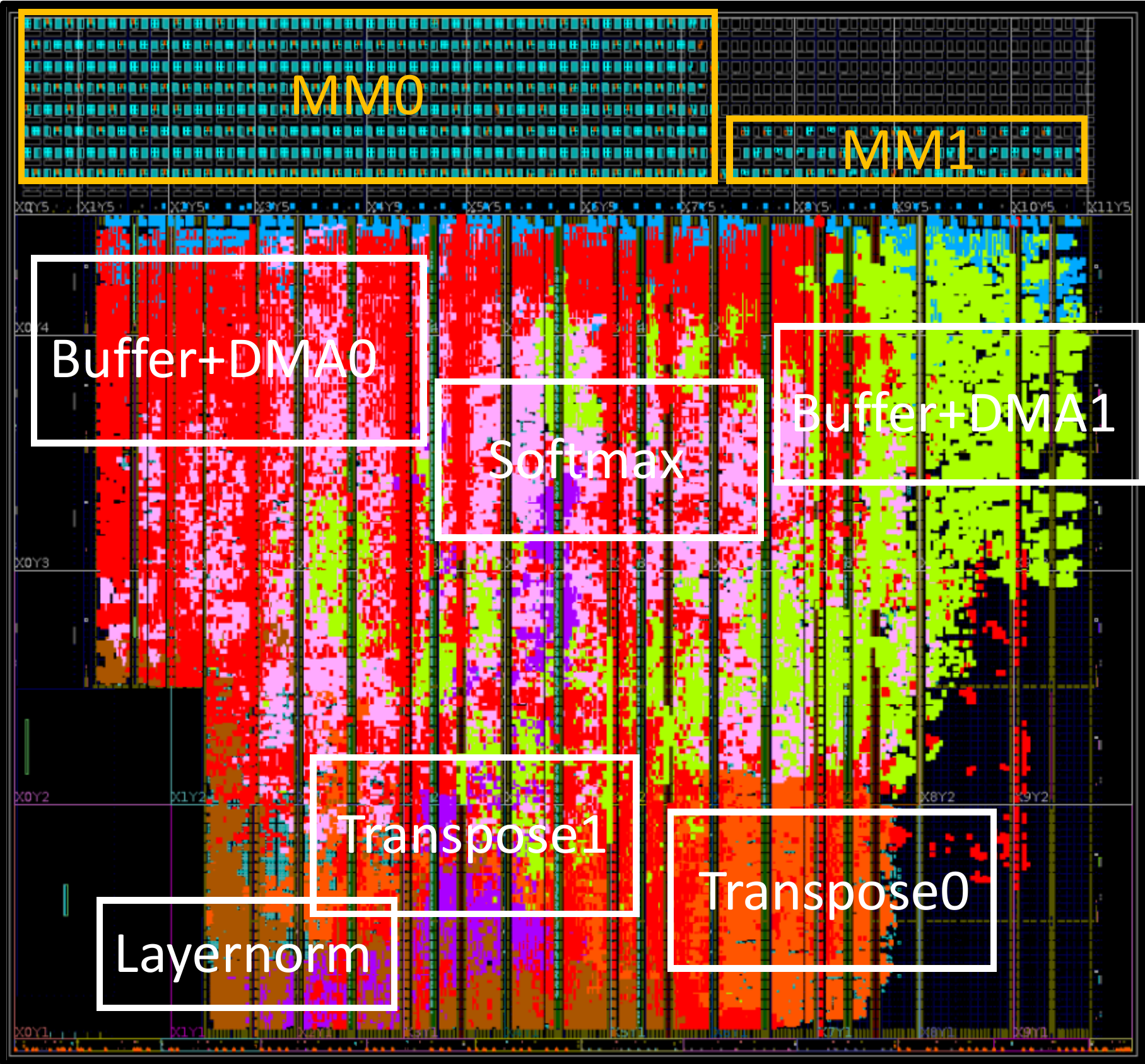}
    \vspace{-5pt}
    \caption{System implementation layout of the two-diverse MM accs and four non-MM accs for BERT.}
    \label{fig:System_layout}
    \vspace{-15pt}
\end{figure}

\begin{table*}[!tb]
\caption{On-board throughput and power comparisons under different MM accs configurations for BERT, ViT, NCF, MLP.}
\vspace{-5pt}
\label{tbl:Application_performance}
\begin{adjustbox}{width=2\columnwidth,center}
\begin{tabular}{c | c | c | c | c | c | c | c | c | c  }
   
\toprule
\textbf{App} & \textbf{CHARM cfg} &  \textbf{LUT} &  \textbf{BRAM} & \textbf{URAM} & \textbf{DSP} & \textbf{AIE}  & \textbf{GFLOPS}  &  \textbf{Power(W)} &  \textbf{GFLOPS/W}   (\textbf{Ratio})  \\ 
\hline

\multirow{4}{*}{BERT}& One\_mono     &103959(11.55\%)  & 764  (79.01\%)  &  384  (82.94\%)  &   165  (8.38\%)   & 384 (96\%)  &276.8   &37.0  &7.48   (1x)    \\     
                     & One\_spe      & 90351(10.04\%)  & 515  (53.26\%)  &  64   (13.82\%)  &   117  (5.95\%)   & 256 (64\%)  &515.4   &32.4  &15.91  (2.13x)    \\ 
                     &\textbf{\emph{Two\_diverse}}   &\textbf{\emph{343774(38.20\%)}}  & \textbf{\emph{534  (55.22\%)}}  &  \textbf{\emph{272  (58.75\%)}}  &   \textbf{\emph{442  (22.46\%)}}  & \textbf{\emph{288 (72\%)}}  &\textbf{\emph{1464.2}}  &\textbf{\emph{40.7 }} &\textbf{\emph{35.98 (4.81x)}}    \\     
                     & 8\_duplicate  &222956(24.78\%)  & 664  (68.67\%)  &  384  (82.94\%)  &   488  (24.80\%)  & 256 (64\%)  &534.2   &34.2  &15.62  (2.09x)    \\
\hline
                     
\multirow{4}{*}{ViT} & One\_mono     &103959(11.55\%)  & 764  (79.01\%)  &  384  (82.94\%)  &   165  (8.38\%)   & 384 (96\%)  &49.5    &32.4  &1.53   (1x)    \\     
                     & One\_spe      & 76661(8.52\%)   & 275  (28.44\%)  &  64   (13.82\%)  &   187  (9.50\%)   & 256 (66\%)  &217.1   &28.0  &7.75   (5.08x)    \\ 
                     & \textbf{\emph{Two\_diverse}}  &\textbf{\emph{240563(26.73\%)}}  & \textbf{\emph{590  (61.01\%)}}  &  \textbf{\emph{320  (69.11\%)}}  &   \textbf{\emph{299  (15.19\%)}}  & \textbf{\emph{264 (72\%)}}  &\textbf{\emph{1609.0}}  &\textbf{\emph{39.6}}  &\textbf{\emph{40.63 (26.60x)}}    \\     
                     & 8\_duplicate  &222956(24.78\%)  & 664  (68.67\%)  &  384  (82.94\%)  &   488  (24.80\%)  & 256 (64\%)  &382.2   &32.8  &11.65  (7.63x)    \\
\hline
                     
\multirow{4}{*}{NCF} & \textbf{\emph{One\_mono}}     &\textbf{\emph{103959(11.55\%)}}  & \textbf{\emph{764  (79.01\%)}}  &  \textbf{\emph{384  (82.94\%)}}  &   \textbf{\emph{165  (8.38\%)}}   & \textbf{\emph{384 (96\%)}}  &\textbf{\emph{1736.0}} &\textbf{\emph{45.2}}  &\textbf{\emph{38.41  (1x)}}    \\     
                     & \textbf{\emph{One\_spe}}      &\textbf{\emph{103959(11.55\%)}}  & \textbf{\emph{764  (79.01\%)}}  &  \textbf{\emph{384  (82.94\%)}}  &   \textbf{\emph{165  (8.38\%)}}   & \textbf{\emph{384 (96\%)}}  &\textbf{\emph{1736.0}} &\textbf{\emph{45.2}}  &\textbf{\emph{38.41  (1.00x)}}    \\
                     & Two\_diverse  &161597(17.96\%)  & 790  (81.70\%)  &  352  (76.03\%)  &   326  (16.57\%)  & 384 (96\%)  &1730.9  &45.1  &38.38  (0.99x)    \\      
                     & 8\_duplicate  &222956(24.78\%)  & 664  (68.67\%)  &  384  (82.94\%)  &   488  (24.80\%)  & 256 (64\%)  &671.0   &35.0  &19.17  (0.50x)    \\
\hline
                    
\multirow{4}{*}{MLP} & \textbf{\emph{One\_mono}}     &\textbf{\emph{103959(11.55\%)}}  & \textbf{\emph{764  (79.01\%)}}  &  \textbf{\emph{384  (82.94\%)}}  &   \textbf{\emph{165  (8.38\%)}}   & \textbf{\emph{384 (96\%)}}  &\textbf{\emph{2936.7}}  &\textbf{\emph{51.4}}  &\textbf{\emph{57.13  (1x)}}    \\     
                     & \textbf{\emph{One\_mono}}     &\textbf{\emph{103959(11.55\%)}}  & \textbf{\emph{764  (79.01\%)}}  &  \textbf{\emph{384  (82.94\%)}}  &   \textbf{\emph{165  (8.38\%)}}   & \textbf{\emph{384 (96\%)}}  &\textbf{\emph{2936.7}}  &\textbf{\emph{51.4}}  &\textbf{\emph{57.13  (1.00x)}}    \\  
                     & Two\_diverse  &148158(16.46\%)  & 919  (95.04\%)  &  448  (96.76\%)  &   344  (17.48\%)  & 384 (96\%)  &2386.1  &48.8  &48.90  (0.86x)    \\      
                     & 8\_duplicate  &222956(24.78\%)  & 664  (68.67\%)  &  384  (82.94\%)  &   488  (24.80\%)  & 256 (64\%)  &696.0   &35.2  &19.77  (0.35x)     \\

\bottomrule
\end{tabular}
\end{adjustbox}
\vspace{-5pt}
\end{table*}

\begin{table*}[!tb]
\caption{Resource utilization for each acc in the design for BERT with two MM diverse accs, four non-MM accs.}
\vspace{-8pt}
\label{tbl:Hardware Utilization}
\begin{adjustbox}{width=1.8\columnwidth,center}
\begin{tabular}{c | c | c | c | c | c | c | c }
   
\toprule
\textbf{Type} & \textbf{REG} &  \textbf{LUTLogic} &  \textbf{LUTMem} &  \textbf{BRAM} & \textbf{URAM}  & \textbf{DSP} & \textbf{AIE} \\ 
\midrule
MM0+DMA0+buffer & 96790  (5.55\%)  &  91034  (10.41\%)  &   835   (0.19\%)  & 515 (53.26\%)  & 256(55.29\%)  &246(12.50\%)   & 256(64\%) \\     
MM1+DMA1+buffer & 62415  (3.58\%)  &  94739  (10.83\%)  &   37668 (8.48\%)  & 19  (1.96\%)   & 16 (3.46\%)   &196( 9.96\%)   & 32 (8\%) \\ 
Layernorm       & 45101  (2.58\%)  &  33939  (3.88\%)   &   4234  (0.95\%)  & 15  (1.55\%)   & 90 (19.44\%)  &129( 6.55\%)   & 0 (0\%) \\     
Softmax         & 34270  (1.96\%)  &  33623  (3.84\%)   &   2854  (0.64\%)  & 243 (25.13\%)  & 0  (0\%)      &151( 7.67\%)   & 0  (0\%) \\
Transpose0      & 14217  (0.81\%)  &  6926   (0.79\%)   &   1097  (0.25\%)  & 15  (1.55\%)   & 0  (0\%)      &94 ( 4.78\%)  & 0  (0\%) \\ 
Transpose1      & 33967  (1.95\%)  &  58510  (6.69\%)   &   32512 (7.32\%)  & 15  (1.55\%)   & 0  (0\%)      &19 ( 0.97\%)  & 0  (0\%) \\
\bottomrule
\end{tabular}
\end{adjustbox}
\vspace{-3pt}
\end{table*}

We apply the CHARM framework to four applications, BERT, ViT, NCF, MLP. All the shapes of the MM kernels in these models are listed in Table \ref{tbl:Evaluation}.
We explore the number of accs from 1 to 8 and showcase the representative CHARM designs, including one monolithic MM acc, one specialized MM acc, two-diverse MM accs, and eight-duplicate MM accs, for each application.
The one monolithic MM design is described in section ~\ref{sec:exp_one_acc}, which stays the same for all four applications.
It is set as the baseline design for comparisons.
All the other MM acc designs are customized for each application and are designed and implemented using the CHARM framework.
All the designs of the same application use the same non-MM kernels.
Table~\ref{tbl:Application_performance} reports the on-board throughput and power consumption under different acc configurations for all the four applications. 

CHARM achieves 1.46 TFLOPs, 1.61 TFLOPs, 1.74 TFLOPs, and 2.94 TFLOPs maximum throughput for the MMs in BERT, ViT, NCF, MLP. 
Table~\ref{tbl:time_breakdown} shows the time breakdown for the MM, the layernorm, the softmax, and the transpose for each end-to-end application.
We highlight the best design(s) for each application in Table~\ref{tbl:Application_performance}.
For BERT and ViT, the two-diverse MM accs designs are the best, whereas for NCF and MLP, one-acc designs are the best. 
This is because BERT and ViT have both large and small MMs whereas MLP only has large MMs.
NCF also has both large and small MMs. However, small MMs consume less than 0.8\% of the total computation, and designs favoring the large MMs stand out as the best.
The eight-duplicate designs are inferior for all the applications due to insufficient data reuse for each acc.

For BERT and ViT, when compared to one monolithic design, the customization of using one specialized acc design for a specific application provided by CHARM gives 2.13$\times$, 5.08$\times$ gain on energy efficiency~(GFLOPs/W), respectively. 
The additional design spaces explored by using more than one-acc, with heterogeneous and diverse shaped accs provided by CHARM framework, give us  
2.25$\times$, 5.24$\times$ extra energy efficiency gains for BERT and ViT, respectively.
\textbf{These gains demonstrate the innovative design methodology of CHARM, i.e., composing heterogeneous accelerators.}

We show the implementation layout of the two-diverse MM acc design, i.e., the best design for BERT, in Figure \ref{fig:System_layout}. 
This is also the layout corresponding to Figure~\ref{System_design} that contains two MM accs and four non-MM communication-bound accs. 
The hardware resource utilization for each acc is reported in Table ~\ref{tbl:Hardware Utilization}. The MM acc 0 provides high data reuse and computation efficiency when calculating large MMs by utilizing 256 AIEs, 53.26\% BRAM, and 55.29\% URAM. 
The MM acc 1 utilizes 32 AIEs, 1.96\% BRAM, and 3.46\% URAM which provides the needed computation and communication without resource over-provisioning for small MMs in BERT.

\begin{figure}
    \centering
    \includegraphics[width=0.85\linewidth]{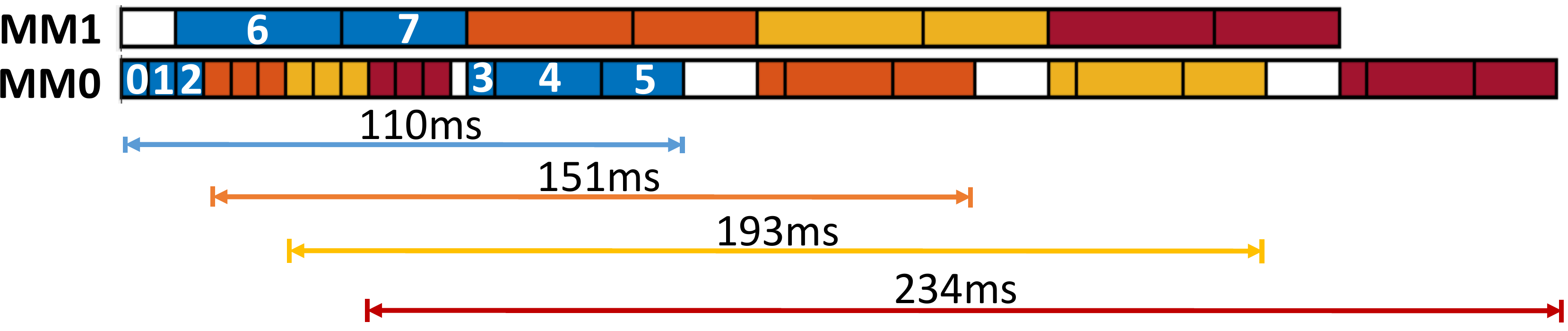}
    \vspace{-10pt}
    \caption{Timeline of 4 tasks scheduled on 2 accs for BERT.}
    \label{fig:Task_scheduler}
    \vspace{-15pt}
\end{figure}

\noindent\textbf{Explore latency-throughput tradeoff in CHARM.}
\label{sec:exp_latency}
As shown in Figure~\ref{fig:Task_scheduler}, we map four concurrent tasks on the BERT design with two-diverse accs.
Each task has eight MM kernels and there are dependency edges, including  
{0$\rightarrow$6, 1$\rightarrow$6, 6$\rightarrow$7, 2$\rightarrow$7, 7$\rightarrow$3$\rightarrow$4$\rightarrow$5}, where x$\rightarrow$y means y depends on x. 
The other non-MM communica\-tion-bound kernels are not shown in the figure for illustrative simplicity.
It takes 110ms to finish the 1st task and 234ms to finish the 4th task.  
For one-acc specialized design, each task latency is 162.6ms.
Therefore, we have a design tradeoff, i.e., one specialized acc design can process fine-grained tasks whereas two-diverse accs design requires coarse-grained tasks to fill the pipeline of the two accs.
Comparing to the one specialized Acc design, with 0.67$\times$, 0.92$\times$, 1.18$\times$, 1.43$\times$ latency for different tasks, we gain 2.8$\times$ overall throughput in return.
This illustrates that the CHARM framework allows explorations on the latency-throughput tradeoff and users can specify targets to let CHARM generate the designs that optimize throughput while meeting the latency requirement or vice versa.

\noindent\textbf{CHARM DSE Efficiency.}
We use CHARM to perform a sort-based two-step search algorithm in CDAC.
For BERT, compared to the exhaustive search, CDAC finds the optimal solution in 170 seconds whereas the exhaustive search takes 33 mins (\#search iterations: 2M vs. 58M) with MATLAB R2021b on an Intel Core i9-10900X CPU.

\section{Discussion of Architecture Insight and Mapping Insight}
\label{sec:insight}

\begin{figure}
    \centering
    \includegraphics[width=0.85\linewidth]{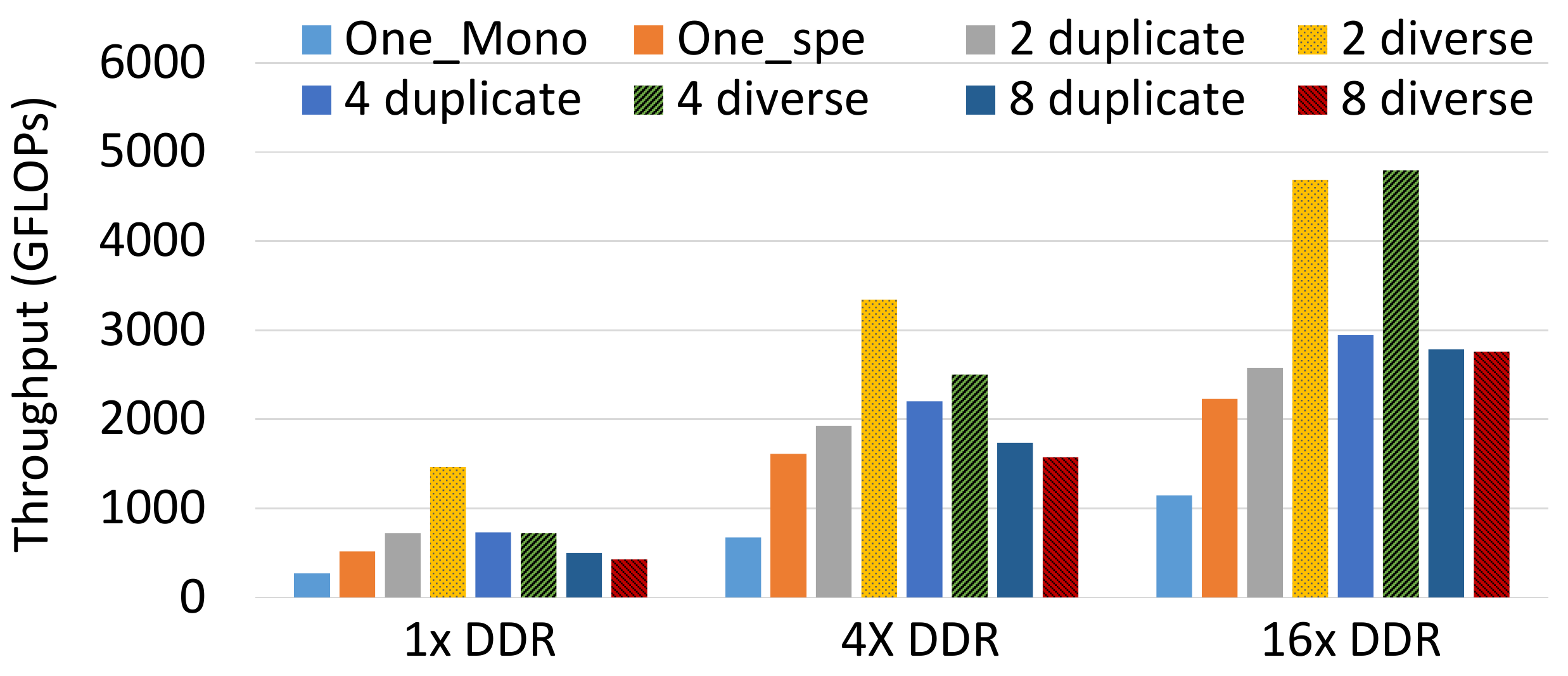}
    \vspace{-15pt}
    \caption{Throughput comparison under different off-chip bandwidth configurations from CHARM for BERT.}
    \vspace{-15pt}
    \label{fig:throughput_ddr}
\end{figure}

By leveraging the strong modeling capability provided by the CHARM framework, we explore performance under different hardware architecture changes (number of AIEs, on-chip storage size, off-chip bandwidth) to do pre-silicon architecture explorations and provide architecture design insights that could be helpful for future generation devices.
Here, we leverage the CHARM modeling to report throughput for different acc configurations including 1-, 2-, 4-, and 8-accs in the system. 
For each design (except one-acc), we have two variants, duplicate accs or diverse accs. 
The explorations help us to understand the following research questions:

\noindent\textbf{Q1: Can we benefit from higher off-chip bandwidth?
}\\
\textcolor{black}{\noindent\textbf{A1: Yes. Versal needs higher off-chip bandwidth.
}}\\
We first explore the performance, assuming the platform has more off-chip bandwidth. 
We increase the DDR bandwidth by 4$\times$ to simulate multiple DDR banks and by 16$\times$ to simulate the case when we have a high bandwidth memory (HBM).
As shown in Figure~\ref{fig:throughput_ddr}, 
the throughput from the best design for BERT in each bandwidth configuration rises from 1.48 TFLOPs to 3.34 TFLOPs with 4$\times$ bandwidth and to 4.80 TFLOPs with 16$\times$ bandwidth.
The improvement from 1$\times$ to 4$\times$ DDR is within expectation and implies that the designs for BERT are bounded by off-chip bandwidth.
The maximum throughput for 16$\times$ is bounded by the system computation throughput as 4.8 TFLOPs, which is constrained by single kernel computation efficiency (95\%) and PL$\leftrightarrow$AIE efficiency(85\%).
Another observation from Figure~\ref{fig:throughput_ddr} is that the throughput improvement of multiple accs is larger than that of the single acc since when the number of accs increases, each acc has less data reuse and tends to be more bounded by the off-chip bandwidth.

\begin{figure}
    \vspace{-5pt}
    \centering
    \includegraphics[width=0.9\linewidth]{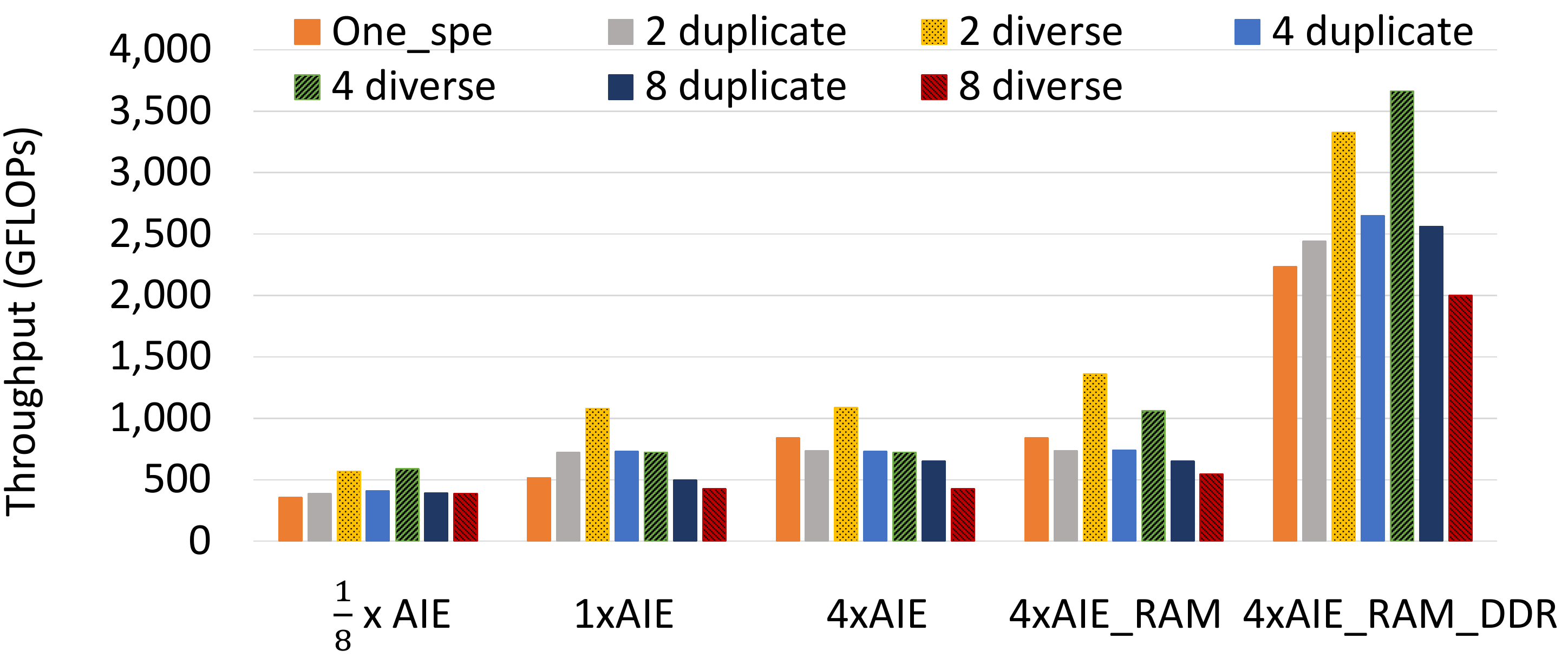}
    \vspace{-15pt}
    \caption{Throughput comparison under different AIE, on-chip, and off-chip  configurations from CHARM for BERT.}
    \vspace{-10pt}
    \label{fig:throughput_aie}
\end{figure}

\noindent\textbf{Q2: Can we leverage CHARM in future architectures?} \\
\noindent\textbf{A2: Yes. The last group in Figure~\ref{fig:throughput_aie} implies that as the computation and communication parallelism further increases in the future, there is a need for more heterogeneous accelerator architectures and CHARM can serve as one of the most promising  solutions. 
}\\
We  explore the performance by varying the number of AIEs, on-chip RAM, and off-chip bandwidth. 
We reduce the number of AIEs to 1/8 of the current AIE array size to simulate the computation capacity of the previous generation FPGA where only PL is equipped with DSPs and has about 1/8 of the theoretical fp32 peak performance of Versal ACAP. 
As shown in the first group in Figure~\ref{fig:throughput_aie},
the performance difference between the minimum and the maximum under different acc configurations is less than 40\%.
As the computation parallelism is reduced to 1/8, the waste resulting from the inconsistency between the massive parallelism and the small MM size is mitigated.
On the other hand, as shown in the last group in Figure~\ref{fig:throughput_aie}, 4-diverse acc stands out as the best when we increase AIE, on-chip storage, and off-chip bandwidth all by 4$\times$.
Simply increasing AIEs does not give significant improvement whereas increasing all the resources as a whole does.


\section{Conclusion and Acknowledgement}
In this paper, we propose the CHARM architecture and the CHARM framework to provide a novel system-level design methodology for composing heterogeneous accelerators for different MMs within an application and generating end-to-end application solutions. 
We will explore and extend CHARM for more applications and more data types in our future work.

We acknowledge the support from the University of Pittsburgh New Faculty Start-up Grant, NSF awards \#2213701, \#2217003 
and the support from CRISP, one of six SRC JUMP centers.
We thank all the reviewers for their valuable feedback and Marci Baun for helping edit the paper. 
We thank AMD/Xilinx for FPGA and software donation, and support from the AMD/Xilinx Center of Excellence at UIUC, the AMD/Xilinx Heterogeneous Accelerated Compute Cluster at UCLA, and the Center for Research Computing (CRC) at University of Pittsburgh.



\bibliographystyle{unsrt}
\bibliography{reference}

\begin{thebibliography}{10}

\bibitem{vaswani2017attention}
Ashish Vaswani, Noam Shazeer, Niki Parmar, Jakob Uszkoreit, Llion Jones,
  Aidan~N Gomez, {\L}ukasz Kaiser, and Illia Polosukhin.
\newblock Attention is all you need.
\newblock {\em Advances in neural information processing systems}, 30, 2017.

\bibitem{he2017neural}
Xiangnan He, Lizi Liao, Hanwang Zhang, Liqiang Nie, Xia Hu, and Tat-Seng Chua.
\newblock Neural collaborative filtering.
\newblock In {\em Proceedings of the 26th international conference on world
  wide web}, pages 173--182, 2017.

\bibitem{dosovitskiy2020image}
Alexey Dosovitskiy, Lucas Beyer, Alexander Kolesnikov, Dirk Weissenborn,
  Xiaohua Zhai, Thomas Unterthiner, Mostafa Dehghani, Matthias Minderer, Georg
  Heigold, Sylvain Gelly, et~al.
\newblock An image is worth 16x16 words: Transformers for image recognition at
  scale.
\newblock {\em arXiv preprint arXiv:2010.11929}, 2020.

\bibitem{wang2019benchmarking}
Yu~Emma Wang, Gu-Yeon Wei, and David Brooks.
\newblock {Benchmarking TPU, GPU, and CPU platforms for deep learning}.
\newblock {\em arXiv preprint arXiv:1907.10701}, 2019.

\bibitem{jouppi2017datacenter}
Norman~P Jouppi, Cliff Young, Nishant Patil, David Patterson, Gaurav Agrawal,
  Raminder Bajwa, Sarah Bates, Suresh Bhatia, Nan Boden, Al~Borchers, et~al.
\newblock In-datacenter performance analysis of a tensor processing unit.
\newblock In {\em Proceedings of the 44th annual international symposium on
  computer architecture}, pages 1--12, 2017.

\bibitem{mutlu2023modern}
Onur Mutlu, Saugata Ghose, Juan G{\'o}mez-Luna, and Rachata Ausavarungnirun.
\newblock A modern primer on processing in memory.
\newblock In {\em Emerging Computing: From Devices to Systems}, pages 171--243.
  Springer, 2023.

\bibitem{oliveira2021damov}
Geraldo~F Oliveira, Juan G{\'o}mez-Luna, Lois Orosa, Saugata Ghose, Nandita
  Vijaykumar, Ivan Fernandez, Mohammad Sadrosadati, and Onur Mutlu.
\newblock {DAMOV: A new methodology and benchmark suite for evaluating data
  movement bottlenecks}.
\newblock {\em IEEE Access}, 9:134457--134502, 2021.

\bibitem{hassan2019crow}
Hasan Hassan, Minesh Patel, Jeremie~S Kim, A~Giray Yaglikci, Nandita
  Vijaykumar, Nika~Mansouri Ghiasi, Saugata Ghose, and Onur Mutlu.
\newblock Crow: A low-cost substrate for improving dram performance, energy
  efficiency, and reliability.
\newblock In {\em Proceedings of the 46th International Symposium on Computer
  Architecture}, pages 129--142, 2019.

\bibitem{demmel2012communication}
Jim Demmel.
\newblock Communication avoiding algorithms.
\newblock In {\em 2012 SC Companion: High Performance Computing, Networking
  Storage and Analysis}, pages 1942--2000. IEEE, 2012.

\bibitem{Versal_ACAP}
{AMD/Xilinx}.
\newblock {Versal Adaptive Compute Acceleration Platform}.

\bibitem{VersalDPU}
{AMD}.
\newblock {IP Overlays of Deep learning Processing Unit }, 2022.

\bibitem{chen2016eyeriss}
Yu-Hsin Chen et~al.
\newblock Eyeriss: A spatial architecture for energy-efficient dataflow for
  convolutional neural networks.
\newblock {\em ACM SIGARCH Computer Architecture News}, 2016.

\bibitem{chen2019eyeriss}
Yu-Hsin Chen, Tien-Ju Yang, Joel Emer, and Vivienne Sze.
\newblock Eyeriss v2: A flexible accelerator for emerging deep neural networks
  on mobile devices.
\newblock {\em IEEE Journal on Emerging and Selected Topics in Circuits and
  Systems}, 9(2):292--308, 2019.

\bibitem{du2015shidiannao}
Zidong Du, Robert Fasthuber, Tianshi Chen, Paolo Ienne, Ling Li, Tao Luo,
  Xiaobing Feng, Yunji Chen, and Olivier Temam.
\newblock Shidiannao: Shifting vision processing closer to the sensor.
\newblock In {\em Proceedings of the 42nd Annual International Symposium on
  Computer Architecture}, pages 92--104, 2015.

\bibitem{nurvitadhi2019compete}
Eriko Nurvitadhi, Dongup Kwon, Ali Jafari, Andrew Boutros, Jaewoong Sim,
  Phillip Tomson, Huseyin Sumbul, Gregory Chen, Phil Knag, Raghavan Kumar,
  et~al.
\newblock Why compete when you can work together: Fpga-asic integration for
  persistent rnns.
\newblock In {\em 2019 IEEE 27th Annual International Symposium on
  Field-Programmable Custom Computing Machines (FCCM)}, pages 199--207. IEEE,
  2019.

\bibitem{boutros2020beyond}
Andrew Boutros, Eriko Nurvitadhi, Rui Ma, Sergey Gribok, Zhipeng Zhao, James~C
  Hoe, Vaughn Betz, and Martin Langhammer.
\newblock Beyond peak performance: Comparing the real performance of
  ai-optimized fpgas and gpus.
\newblock In {\em 2020 International Conference on Field-Programmable
  Technology (ICFPT)}, pages 10--19. IEEE, 2020.

\bibitem{fowers2018configurable}
Jeremy Fowers, Kalin Ovtcharov, Michael Papamichael, Todd Massengill, Ming Liu,
  Daniel Lo, Shlomi Alkalay, Michael Haselman, Logan Adams, Mahdi Ghandi,
  et~al.
\newblock A configurable cloud-scale dnn processor for real-time ai.
\newblock In {\em 2018 ACM/IEEE 45th Annual International Symposium on Computer
  Architecture (ISCA)}, pages 1--14. IEEE, 2018.

\bibitem{de2020fblas}
Tiziano De~Matteis, Johannes de~Fine~Licht, and Torsten Hoefler.
\newblock {FBLAS: Streaming linear algebra on FPGA}.
\newblock In {\em SC20: International Conference for High Performance
  Computing, Networking, Storage and Analysis}, pages 1--13. IEEE, 2020.

\bibitem{de2020flexible}
Johannes de~Fine~Licht, Grzegorz Kwasniewski, and Torsten Hoefler.
\newblock Flexible communication avoiding matrix multiplication on fpga with
  high-level synthesis.
\newblock In {\em Proceedings of the 2020 ACM/SIGDA International Symposium on
  Field-Programmable Gate Arrays}, pages 244--254, 2020.

\bibitem{zhang2015optimizing}
Chen Zhang et~al.
\newblock Optimizing fpga-based accelerator design for deep convolutional
  neural networks.
\newblock In {\em Proc. of FPGA}, pages 161--170. ACM, 2015.

\bibitem{Moss_Krishnan_Nurvitadhi_Ratuszniak_Johnson_Sim_Mishra_Marr_Subhaschandra_Leong_2018}
Duncan J.~M. Moss, Srivatsan Krishnan, Eriko Nurvitadhi, Piotr Ratuszniak,
  Chris Johnson, Jaewoong Sim, Asit Mishra, Debbie Marr, Suchit Subhaschandra,
  and Philip H.~W. Leong.
\newblock A customizable matrix multiplication framework for the intel harpv2
  xeon+fpga platform: A deep learning case study.
\newblock In {\em Proceedings of the 2018 ACM/SIGDA International Symposium on
  Field-Programmable Gate Arrays}, FPGA ’18, page 107–116. Association for
  Computing Machinery, Feb 2018.

\bibitem{Wang_Guo_Cong_2021}
Jie Wang, Licheng Guo, and Jason Cong.
\newblock {AutoSA: A Polyhedral Compiler for High-Performance Systolic Arrays
  on FPGA}.
\newblock In {\em The 2021 ACM/SIGDA International Symposium on
  Field-Programmable Gate Arrays}, FPGA ’21, page 93–104. Association for
  Computing Machinery, Feb 2021.

\bibitem{song2022sextans}
Linghao Song, Yuze Chi, Atefeh Sohrabizadeh, Young-kyu Choi, Jason Lau, and
  Jason Cong.
\newblock Sextans: A streaming accelerator for general-purpose sparse-matrix
  dense-matrix multiplication.
\newblock In {\em Proceedings of the 2022 ACM/SIGDA International Symposium on
  Field-Programmable Gate Arrays}, FPGA '22, page 65–77, New York, NY, USA,
  2022. Association for Computing Machinery.

\bibitem{song2022serpens}
Linghao Song, Yuze Chi, Licheng Guo, and Jason Cong.
\newblock Serpens: A high bandwidth memory based accelerator for
  general-purpose sparse matrix-vector multiplication.
\newblock In {\em Proceedings of the 59th ACM/IEEE Design Automation
  Conference}, pages 211--216, 2022.

\bibitem{fccm18latte}
Jason Cong, Peng Wei, Cody~Hao Yu, and Peipei Zhou.
\newblock {Latte: Locality Aware Transformation for High-Level Synthesis}.
\newblock In {\em 2018 IEEE 26th Annual International Symposium on
  Field-Programmable Custom Computing Machines (FCCM)}, pages 125--128, 2018.

\bibitem{fccm16}
Peipei Zhou, Hyunseok Park, Zhenman Fang, Jason Cong, and André DeHon.
\newblock {Energy Efficiency of Full Pipelining: A Case Study for Matrix
  Multiplication}.
\newblock In {\em 2016 IEEE 24th Annual International Symposium on
  Field-Programmable Custom Computing Machines (FCCM)}, pages 172--175, 2016.

\bibitem{fpga21mocha}
Peipei Zhou, Jiayi Sheng, Cody~Hao Yu, Peng Wei, Jie Wang, Di~Wu, and Jason
  Cong.
\newblock {MOCHA: Multinode Cost Optimization in Heterogeneous Clouds with
  Accelerators}.
\newblock In {\em The 2021 ACM/SIGDA International Symposium on
  Field-Programmable Gate Arrays}, FPGA '21, page 273–279, New York, NY, USA,
  2021. Association for Computing Machinery.

\bibitem{zhang2018dnnbuilder}
Xiaofan Zhang et~al.
\newblock Dnnbuilder: an automated tool for building high-performance dnn
  hardware accelerators for fpgas.
\newblock In {\em Proc. ICCAD}, page~56. ACM, 2018.

\bibitem{zhang2020dnnexplorer}
Xiaofan Zhang, Hanchen Ye, Junsong Wang, Yonghua Lin, Jinjun Xiong, Wen-mei
  Hwu, and Deming Chen.
\newblock {DNNExplorer: a framework for modeling and exploring a novel paradigm
  of FPGA-based DNN accelerator}.
\newblock In {\em Proceedings of the 39th International Conference on
  Computer-Aided Design}, pages 1--9, 2020.

\bibitem{gao2017tetris}
Mingyu Gao, Jing Pu, Xuan Yang, Mark Horowitz, and Christos Kozyrakis.
\newblock Tetris: Scalable and efficient neural network acceleration with 3d
  memory.
\newblock In {\em Proceedings of the Twenty-Second International Conference on
  Architectural Support for Programming Languages and Operating Systems}, pages
  751--764, 2017.

\bibitem{gao2019tangram}
Mingyu Gao, Xuan Yang, Jing Pu, Mark Horowitz, and Christos Kozyrakis.
\newblock Tangram: Optimized coarse-grained dataflow for scalable nn
  accelerators.
\newblock In {\em Proceedings of the Twenty-Fourth International Conference on
  Architectural Support for Programming Languages and Operating Systems}, pages
  807--820, 2019.

\bibitem{kwon2021heterogeneous}
Hyoukjun Kwon, Liangzhen Lai, Michael Pellauer, Tushar Krishna, Yu-Hsin Chen,
  and Vikas Chandra.
\newblock Heterogeneous dataflow accelerators for multi-dnn workloads.
\newblock In {\em 2021 IEEE International Symposium on High-Performance
  Computer Architecture (HPCA)}, pages 71--83. IEEE, 2021.

\bibitem{NVIDIA}
Nvidia.
\newblock Website.
\newblock \url{http://nvdla.org/}.

\bibitem{FPCA14fccm}
Jason Cong, Hui Huang, Chiyuan Ma, Bingjun Xiao, and Peipei Zhou.
\newblock {A Fully Pipelined and Dynamically Composable Architecture of CGRA}.
\newblock In {\em 2014 IEEE 22nd Annual International Symposium on
  Field-Programmable Custom Computing Machines}, pages 9--16, 2014.

\bibitem{charm12islped}
Jason Cong, Mohammad~Ali Ghodrat, Michael Gill, Beayna Grigorian, and Glenn
  Reinman.
\newblock {CHARM: A Composable Heterogeneous Accelerator-Rich Microprocessor}.
\newblock In {\em Proceedings of the 2012 ACM/IEEE International Symposium on
  Low Power Electronics and Design}, ISLPED '12, page 379–384, New York, NY,
  USA, 2012. Association for Computing Machinery.

\bibitem{vck190}
{AMD/Xilinx}.
\newblock {Versal AI Core Series VCK190 Evaluation Kit}, 2022.

\bibitem{aiearch}
{AMD/Xilinx}.
\newblock {AI Engine Technology}, 2022.

\bibitem{cong2011high}
Jason Cong, Bin Liu, Stephen Neuendorffer, Juanjo Noguera, Kees Vissers, and
  Zhiru Zhang.
\newblock {High-level synthesis for FPGAs: From prototyping to deployment}.
\newblock {\em IEEE Transactions on Computer-Aided Design of Integrated
  Circuits and Systems}, 30(4):473--491, 2011.

\bibitem{cong2022fpga}
Jason Cong, Jason Lau, Gai Liu, Stephen Neuendorffer, Peichen Pan, Kees
  Vissers, and Zhiru Zhang.
\newblock {FPGA HLS Today}: successes, challenges, and opportunities.
\newblock {\em ACM Transactions on Reconfigurable Technology and Systems
  (TRETS)}, 15(4):1--42, 2022.

\bibitem{papakonstantinou2009fcuda}
Alexandros Papakonstantinou, Karthik Gururaj, John~A Stratton, Deming Chen,
  Jason Cong, and Wen-Mei~W Hwu.
\newblock {FCUDA: Enabling efficient compilation of CUDA kernels onto FPGAs}.
\newblock In {\em 2009 IEEE 7th Symposium on Application Specific Processors},
  pages 35--42. IEEE, 2009.

\bibitem{papakonstantinou2011multilevel}
Alexandros Papakonstantinou, Yun Liang, John~A Stratton, Karthik Gururaj,
  Deming Chen, Wen-Mei~W Hwu, and Jason Cong.
\newblock Multilevel granularity parallelism synthesis on fpgas.
\newblock In {\em 2011 IEEE 19th Annual International Symposium on
  Field-Programmable Custom Computing Machines}, pages 178--185. IEEE, 2011.

\bibitem{liang2012high}
Yun Liang, Kyle Rupnow, Yinan Li, Dongbo Min, Minh~N Do, and Deming Chen.
\newblock High-level synthesis: productivity, performance, and software
  constraints.
\newblock {\em Journal of Electrical and Computer Engineering}, 2012, 2012.

\bibitem{tapa21}
Yuze Chi, Licheng Guo, Jason Lau, Young-kyu Choi, Jie Wang, and Jason Cong.
\newblock Extending high-level synthesis for task-parallel programs.
\newblock In {\em 2021 IEEE 29th Annual International Symposium on
  Field-Programmable Custom Computing Machines (FCCM)}, pages 204--213, 2021.

\bibitem{ADF}
{AMD/Xilinx}.
\newblock {Adaptive Data Flow API}.

\bibitem{BEAM}
{AMD/Xilinx}.
\newblock {Board evaluation and management Tool}.

\bibitem{AIE_Intrinsics}
{AMD/Xilinx}.
\newblock {AI Engine API and Intrinsics User Guide}.

\bibitem{AIE_simulator}
{AMD/Xilinx}.
\newblock {Versal™ ACAP AI Engine System C simulator}.

\bibitem{zhang2022h}
Chengming Zhang, Tong Geng, Anqi Guo, Jiannan Tian, Martin Herbordt, Ang Li,
  and Dingwen Tao.
\newblock {H-GCN: A graph convolutional network accelerator on versal acap
  architecture}.
\newblock {\em arXiv preprint arXiv:2206.13734}, 2022.

\end{thebibliography}

\end{document}